\documentclass[aps,pra,reprint]{revtex4-1}

\usepackage{amsmath,amssymb,graphicx,color}

\newcommand{\eqn}{\begin{equation}}
\newcommand{\nqe}{\end{equation}}

\newcommand{\bra}[1]{\langle #1 \rvert}
\newcommand{\ket}[1]{\lvert #1 \rangle}

\newcommand{\abs}[1]{\lvert #1 \rvert}
\DeclareMathOperator{\rmd}{d}
\DeclareMathOperator{\sgn}{sgn}

\DeclareMathOperator{\erfc}{erfc}

\numberwithin{equation}{section}

\begin{document}

\title{An exact formalism for the quench dynamics of integrable models} \author{Deepak
  Iyer} \author{Huijie Guan} \author{Natan Andrei}
\affiliation{Department of Physics and
  Astronomy\\ Rutgers University\\
  Piscataway, New Jersey 08854.}  \date{\today}

\begin{abstract}
  We describe a formulation for studying the quench dynamics of
  integrable systems generalizing an approach by Yudson.  We study the
  evolution of the Lieb-Liniger model, a gas of interacting bosons
  moving on the continuous infinite line and interacting via a short
  range potential. The formalism allows us to quench the system from
  any initial state. We find that for any value of repulsive coupling
  independently of the initial state the system asymptotes towards a
  strongly repulsive gas, while for any value of attractive coupling,
  the system forms a maximal bound state that dominates at longer
  times.  In either case the system equilibrates but does not
  thermalize.  We
  compare this to quenches in a Bose-Hubbard lattice and show that
  there, initial states determine long-time dynamics independent of
  the sign of the coupling.
\end{abstract}

\maketitle

\section{Introduction}
\label{sec:intro}

Nonequilibrium processes occur in fields as diverse as metallurgy and
cell biology -- in fact, most physical processes that occur in nature
are dynamical. It is therefore important to understand nonequilibrium
phenomena from a variety of perspectives and in several different
contexts.

In the context of physics, the dynamics of nonequilibrium processes
has been studied since the early days of thermodynamics (see
Ref.~\onlinecite{grootbk} for a historical introduction to the
subject).  More recently, the study of quantum nonequilibrium
processes has received a boost from highly tunable experimental
systems, in particular ultracold atomic gases, that can be well
isolated from their environment. They provide a testing ground for
theories of quantum nonequilibrium behavior and have led to
observations of interesting new phenomena~\cite{blochdali}.

Unlike thermodynamics, there exists no general framework to understand
far from equilibrium behavior.  Different problems need different
approaches and in most cases we can only understand the behavior of
physical observables in limited regimes of parameters or windows of
time.  However, a good sense of the complexity of the problem is
emerging from a multitude of studies on several types of systems
analytically, computationally, and experimentally.

Within the domain of quantum phenomena, transport and quenches are the
primary means of studying nonequilibrium physics. Transport phenomena
include for example, transient and steady-state currents in devices,
study of quantum impurity models and their conduction properties,
quantum hall edge states, and the surface states of topological
insulators.  Quenches, on the other hand, allow us to study phenomena
like thermalization and relaxation of physical observables by strongly
disturbing an equilibrium system and watching it evolve~\footnote{In
  the broader sense transport may be viewed as the long time limit of
  quench. For example, a steady state current would result if we
  quench a quantum dot attaching it to two leads held at different
  chemical potentials~\cite{doyon}}. Cold atom systems, as noted, are
particularly amenable to quenches -- we briefly describe these in
section~\ref{sec:expt}. On the theoretical side, with advances in
computational techniques, it is possible to calculate the time
evolution of various observables of larger systems and study the
underlying physics~\cite{rigol1,rigol2,calabrese,caux,white,daley}. However, these
techniques are not suitable for very large systems or continuum
models.

In this article, we are interested in studying the quench dynamics of
systems described by integrable models. These models possess an
infinite number of conserved quantities, and this property is expected
to reveal itself in their dynamics~\cite{rigol1,rigol2,kinoshita}. A
generic model would access all of the constant energy surface in phase
space as it evolves, but this may not be the case for an integrable
model. As we will mention later
in more detail, this work studies the expansion of a local initial state
into infinite volume, and therefore is different from studying the
thermodynamics. The approach we take here mimics experiments directly.

The eigenstates of integrable models can be obtained via the
\emph{Bethe Ansatz} technique~\cite{bethe}, which allows the construction
of a complete set of eigenstates of the Hamiltonian. The Bethe Ansatz has
proved to be of tremendous use in studying the ground states and
thermodynamics of such models~\cite{takahashibk, andrei, tsvelik,
  andreitrieste}.  However, in spite of knowing the eigenstates of the
Hamiltonian the dynamics still remains a complicated problem. In the
remainder of this paper, we elucidate a framework for dynamics
introduced by Yudson~\cite{yudson1,yudson2}. We introduce some
generalizations and use it to understand the quench dynamics of a gas
of bosons with attractive or repulsive short range interactions in the
context of the Lieb-Liniger model~\cite{lieblin}.  This approach
allows analytical calculations that are essential for a complete
understanding of the out-of-equilibrium properties of the system.

The article is a follow up to Ref.~\onlinecite{deeprl}. Some of the
formulas are derived in detail, and some of the plots are repeated for
completeness. We also correct an error in the earlier work.

Before getting into details we discuss some general
issues concerning the quench dynamics of isolated systems defined on
the infinite line.

\subsection{Dynamics of $1d$ isolated many-body systems}
\label{sec:opensys}

While studying the thermodynamic properties of a quantum system, one
needs to enumerate and classify the eigenstates of the Hamiltonian in
order to construct the partition function. To achieve this, some
finite volume boundary conditions (BC) are typically imposed ---
either periodic BC to maintain translation invariance or open BC when
the system has physical ends. One can then identify the ground state
and the low lying excitations that dominate the low-temperature
physics. In the \emph{thermodynamic limit}
$\frac{N\to\infty}{L\to\infty} \equiv \rho$, the limit of very large
systems with large number of particles at finite density, the effect
of the boundary condition is negligible and we expect the results to
be universally valid.

When the system is out of equilibrium, a different set of issues
arises.  We shall consider here the process of a ``quantum quench''
where one studies the time evolution of a system after a sudden
change in the parameters of the Hamiltonian governing the system. To
be precise, one assumes that the system starts in some stationary
state $\ket{\Psi_0}$. This stationary state can be thought of as the
ground state of some (interacting) Hamiltonian $H_0$. Following the
quench at $t=0$, the system evolves in time under the influence of a
new Hamiltonian $H$ which may differ from $H_{0}$ in many ways. One
may add an interaction, change an interaction coupling constant, apply
or remove an external potential or increase the size of the system.
Further, the quench can be sudden, i.e., over a time window much
shorter than other time scales in the system, driven at a constant
rate or with a time dependent ramp.

\begin{figure}[htb]
  \centering \includegraphics[width=8cm]{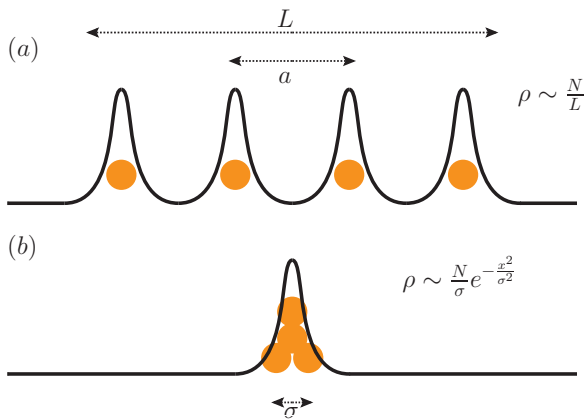}
  \caption{Initial states. $(a)$ For $\frac{a}{\sigma}\gg 1$, we have
    a lattice like state, $\ket{\Psi_{\text{latt}}}$. $(b)$ For $a=0$,
    we have a condensate like state $\ket{\Psi_{\text{cond}}}$,
    $\sigma$ determines the spread.}
  \label{fig:initstate}
\end{figure}

In this paper we shall concentrate on sudden quenches where the
initial state $\ket{\Psi_{0}}$ describes a system in a finite region
of space with a particular density profile: lattice-like, or
condensate-like (see Fig.~\ref{fig:initstate}). Under the effect of
the quenched Hamiltonian the system evolves as $\ket{\Psi_{0}, t}=
e^{-iHt} \ket{\Psi_{0}}$.

To compute the evolution, it is convenient to expand the initial state
in the eigenbasis of the evolution Hamiltonian,
\begin{equation}
  \ket{\Psi_{0}} = \sum_{\{n\}}C_{n}\ket{n},
\end{equation}
where $\ket{n}$ are the eigenstates of $H$ and $C_{n}=
\bra{n}\Psi_{0}\rangle$ are the overlaps with the initial state,
determining the weights with which different eigenstates contribute to
the time evolution:
\begin{equation}\label{eq:quenchoverlaps}
  \ket{\Psi_{0}, t}= \sum_{\{n\}}   e^{-i\epsilon_n t }C_{n}\ket{n}.
\end{equation}
The evolution of observables is then given by,
\begin{equation}
  \begin{split}
    \langle\hat{O}(t)\rangle_{\Psi_0} &= \langle \Psi_{0}, t |\hat{O}\ket{\Psi_{0}, t}\\
    &= \sum_{\{n, m \}} e^{-i(\epsilon_m-\epsilon_n)t}C^{*}_{n}\,C_m
    \langle n |\hat{O} \ket{m},
  \end{split}
\end{equation}
with observables that may be local operators, correlation functions,
currents or global quantities such as entanglements.

\begin{figure}[htb]
  \centering \includegraphics[width=1.5in]{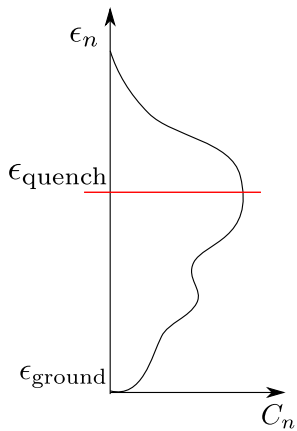}
  \caption{Difference between quench dynamics and
    thermodynamics. After a quench, the system probes high energy
    states and does not necessarily relax to the ground state. In
    thermodynamics, we minimize the energy (or free energy) of a
    system and probe the region near the ground state.}
  \label{fig:energyscales}
\end{figure}

The time evolution is characterized by the energy of the initial
state,
\begin{equation}
  \epsilon_{\text{quench}} = \bra{\Psi_{0}}H\ket{\Psi_{0}} = \sum_{\{n\}} \epsilon_{n}\abs{C_{n}}^{2}
\end{equation}
which is conserved throughout the evolution, specifying the
\emph{energy surface} on which the system moves. This surface is
determined by the initial state through the overlaps $C_n$.  Unlike
the situation in thermodynamics where the ground state and low-lying
excitations play a central role, this is not the case
out-of-equilibrium. A quench puts energy into the system which the
isolated system cannot dissipate and it cannot relax to its ground
state. Rather, the eigenstates that contribute to the dynamics depend
strongly on the initial state via the overlaps $C_n$ (see
Fig.~\ref{fig:energyscales}).

A vivid illustration comes from comparing quenches in systems that
differ in the sign of the interaction. In the Bose-Hubbard model and
the XXZ model it has been observed that the sign of the interaction
plays no role in the quench dynamics~\cite{lahinietal,barmettler},
even though the ground states that correspond to the different signs
are very different. For example, for the XXZ magnet, the ground state
is either ferromagnetic or N\'eel ordered (RVB in 1$d$) depending on
the sign of the anisotropy $\Delta$. In Appendix \ref{sec:bhtebd}, we
show results for the Bose-Hubbard model, and provide an argument for
this observation.  The Lieb-Liniger model, whose quench dynamics we
describe here, on the other hand shows very different behavior,
reaching an long time equilibrium state that depends mainly on the
sign of the interaction.

In the experiments that we seek to describe, a system of $N$ bosons is
initially confined to a region of space of size $L$ and then allowed
to evolve on the infinite line while interacting with short range
interactions.  It is important to first understand the time-scales of
the phenomena that we are studying. There are two main types of time
scales here. One is determined by the initial condition (spatial
extent, overlap of nearby wave-functions), and the other by the
parameters of the quenched system (mass, interaction strength).

For an extended system where we start with a locally uniform density
(see Fig.~\ref{fig:initstate}$a$), we expect the dynamics to be in the
constant density regime as long as $t \ll \frac{L}{v}$, $v$ being the
characteristic velocity of propagation. Although the low energy
thermodynamics of a constant density Bose gas can be described by a
Luttinger liquid~\cite{giamarchibk}, we expect the collective
excitations of the quenched system to behave as a highly excited
Liquid since the initial state is far from the ground state. It is
also possible that depending on the energy density
$\epsilon_{\text{quench}}/L$, the Luttinger liquid description may
break down altogether.

\begin{figure}[htb]
  \centering \includegraphics[width=8cm]{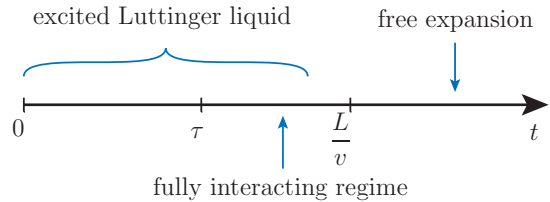}
  \caption{Time scales involved in quench dynamics. $\tau$ is an
    intrinsic time scale that depends on the interaction
    strength. $L/v$ is a characteristic time at which the system sees
    the finite extent.}
  \label{fig:timescales}
\end{figure}

The other time scale that enters the description of nonequilibrium
dynamics is the interaction time scale, $\tau$, a measure of the time
it takes the interactions to develop fully: $\tau \sim
\frac{1}{c^{2}}$ for the Lieb-Liniger model~\footnote{One can refine
  the estimate for the interaction time setting $\tau \sim
  \frac{1}{\delta E}$, with $\delta E = \displaystyle{\bra{\Phi_0} H_I
    \ket{\Phi_0}}$.  Also if we start from a lattice-like state,
  $\tau$ will include a short time scale $\tau_{a}\sim\frac{a}{v}$
  before which the system only expands as a non-interacting gas, until
  neighboring wave-functions overlap sufficiently.}.  Assuming $L$ is
large enough so that $\tau \ll \frac{L}{v}$, we expect a fully
interacting regime to be operative at times beyond the interactions
scale until $t \gg \frac{L}{v}$ and the density of the system can no
longer be considered constant, diminishing with time as the system
expands. In the Lieb-Liniger model, this leads to an effective
increase in the coupling constant which manifests itself as
fermionization for repulsive interaction and bound-state correlations
in the case of attractive interactions.  Thus the main operation of
the interaction occurs in the time range $\tau \lesssim t \lesssim
\frac{L}{v}$, over which the wave function rearranges and after which
the system is dilute and freely expands. In this low density limit,
we cannot make contact with thermodynamic ensembles, in particular, the Generalized
Gibbs ensemble~\cite{rigol3,calabresecardy,cauxessler,cauxkonik}.
For the case $L=\infty$, free
expansion is not present. Figure~\ref{fig:timescales} summarizes the
different time-scales involved in a dynamical situation.

We shall also consider initial conditions where the bosons are
``condensed in space'', occupying the same single particle state
characterized by some scale $\sigma$ (see
Fig.~\ref{fig:initstate}$b$).  In this case the short time dynamics is
not present, the time scales at which we can measure the system are
typically much larger than $\frac{\sigma}{N}$ and we expect the
dynamics to be in the strongly interacting and expanding regime.

\subsection{Quench dynamics and the Bethe Ansatz}
\label{sec:intmod}

To carry out the computation of the quench dynamics we need to know
the eigenstates of the propagating Hamiltonian. The Bethe Ansatz
approach is helpful in this respect as it provides us with the
eigenstates of a large class of interacting one dimensional
Hamiltonians.  Many of the Hamiltonians that can be thus be solved are
of fundamental importance in condensed matter physics and have been
proposed to study various experimental situations. A partial list
includes the Heisenberg chain (magnetism), the Hubbard model
(strong-correlations), the Lieb-Liniger model (cold atoms in optical
traps), the Kondo model, and the Anderson model (impurities in metals,
quantum dots)~\cite{bethe, liebwu, lieblin, andrei, andreitrieste,
  tsvelik}.  For a Hamiltonian to possess eigenstates that are given
in the form of a Bethe Ansatz it must have the property that
multi-particle interactions can be consistently factorized into series
of two particle interactions, all of them being
equivalent\footnote{This set of conditions is known as the Yang-Baxter
  equation.}.

While originally formulated to understand the Heisenberg
chain~\cite{bethe}, the technology has been studied extensively and
recast into a more sophisticated algebraic
formulation~\cite{sklyanin,korepinbk}. The usual focus of the Bethe
Ansatz approach has been on the thermodynamic properties of the
system: determining the spectrum, the free energy, and
susceptibilities. Also, considerable efforts were made to compute
correlation functions~\cite{thacker,korepinbk}. In particular, the
Algebraic Bethe Ansatz has been used in conjunction with numerical
methods to calculate so-called form factors which allows access to the
dynamical structure functions~\cite{caux,calabrese,gritsev}.

Using the Bethe Ansatz to extract dynamics one encounters a triply
complicated problem --- the first being to obtain the full spectrum of
the Hamiltonian, the second to calculate overlaps, and the third to
carry out the sum, which in some cases involves sums over large sets
of different ``configurations'' of states.  The overlaps are
particularly difficult to evaluate due to the complicated nature of
the  Bethe eigenstates and their normalization. The problem is more
pronounced in the far from equilibrium case of a quench when the state
we start with suddenly finds itself far away from the eigenstates of
the new Hamiltonian [see eq.~\eqref{eq:quenchoverlaps}], and all the
eigenstates have non-trivial weights in the time-evolution.  In all
but the simplest cases, the problem is non-perturbative and the
existing analytical techniques are not suited for a direct application
to such a situation.

Often quench calculations are carried out in finite volume and the
infinite volume limit is taken at the end of the calculation. This may
be necessary for thermodynamic calculations, as noted, but not for
quench dynamics (see sec.~\ref{sec:opensys}).  We shall instead carry
out the quench directly in the infinite volume limit in which case the
Yudson representation allows us to carry out the calculations in an
efficient way, doing away with some of the difficulties mentioned
above by not requiring any information about the spectrum, and by
using integration as opposed to discrete summation.

More explicitly, the Schrodinger equation for $N$ bosons
$H\ket{\vec{\lambda}}=\epsilon(\vec{\lambda})\ket{\vec{\lambda}}$ is
satisfied for any value of the momenta $\{ \lambda_{j}, j=1,\cdots,N
\}$ if no boundary conditions are imposed.  The initial state can then
be written as $ \ket{\Psi_{0}} = \int d^N \lambda \; C_{\vec{\lambda}}
\; \ket{\vec{\lambda}}$, with the integration over $\vec{\lambda}$
replacing the summation. This is akin to summing over an over-complete
basis, the relevant elements in the sum being automatically picked up
by the overlap with the physical initial state.

It is important to note, however, that while the spectrum of an
infinite system is continuous, it can still be very complicated. This
is indeed the case with several integrable models, including the
Lieb-Liniger model, where the analytic structure of the $S$-matrix
(the momentum dependent phase picked up when two particles cross) may
permit momenta in the form of complex-conjugate pairs signifying bound
states.  In the formalism we employ, such states are taken into
account by appropriately choosing the contours of integration in the
complex plane. This procedure is described in detail.

The remainder of this article is organized as follows. In section
\ref{sec:expt}, we briefly discuss experiments with cold atoms.  In
sections \ref{sec:model} and \ref{sec:yudrep} we will describe the
Lieb-Liniger model and Yudson representation for this model, and show
how we can use it to calculate the time evolution of an arbitrary
initial state. Section \ref{sec:2pquench} studies the two particle
case in detail.  We then go on to calculate the evolution of the noise
correlations for both the repulsive and the attractive gas at long
times in sec.~\ref{sec:asymp}. We conclude with a conjecture about the
nature of equilibration and thermalization and end with a description
of ongoing work and future directions in sec.~\ref{sec:concl}.
Appendix~\ref{sec:bhtebd} discusses a quench in the Bose-Hubbard
model, and why the sign of the interaction doesn't affect the quench
dynamics.

\section{Experiments with cold atoms}
\label{sec:expt}

The experimental arena to which our calculations apply are ultracold
atomic gases in laser traps, which have become a powerful system for
exploring nonequilibrium phenomena~\cite{blochdali,cazalilla}.  The
systems are formed by trapping a gas of atoms using standing light
waves made by lasers. The gases are cooled evaporatively and are well
isolated from any thermal baths making them ideal for studying
relaxation and thermalization in isolated quantum systems. The
interactions between the particles, the potentials, and their
statistics can be controlled by the use of external magnetic and
electric fields, tuning the optical lattice, and loading different
atoms into the traps. Systems with mobile impurities can also be
studied by loading two or more different species of atoms into the
lattices.  Lattices can be three dimensional or can be made quasi-1$d$
or 2$d$ by using confining potentials. The typical relaxation and
evaporation time scales in these systems are in the milliseconds. This
makes measurement easier than in solid state systems. It also allows
for sudden quenches.  Disorder is also largely absent, unless
introduced.

Tuning the parameters allows the study of superfluid behavior, Mott
insulators, spin chains and so on.  Such a gas trapped by lasers and
cooled to nano-Kelvin temperatures can be quenched by suddenly
changing the interaction between the molecules and the external
trapping potential. Evolution can be globally observed by imaging the
gas, and the time evolution of densities and correlation functions can
be obtained from these
images~\cite{blochdali,gorlitz,kinoshita,moritz}.

In one dimension, which is of particular interested to us, the typical
models that are used to study these systems are the Bose-Hubbard
model, the XXZ model, the Sine-Gordon model and the Lieb-Liniger
model. Each of these models studies a different regime of the gas. The
Bose-Hubbard model is optimal for atoms hopping on a one dimensional
lattice. A particular limit of the Bose-Hubbard model can be mapped to
the XXZ spin chain~\cite{barmettler} which is integrable. The
continuum gas is captured by the Lieb-Liniger model.

In this article, we will study the long time dynamics of the
Lieb-Liniger model, also an integrable model as mentioned.  In this
context, it is an important question as to how ``integrable'' a
particular experimental realization is.  Often the experimental setup
maintains an external trapping potential and including such a
potential in an integrable Hamiltonian may render the system
non-integrable. Experimentally, such potentials need to be eliminated
to the extent possible. This can be partially achieved by using
blue/red detuned lasers to create a flat potential well.  As has been
shown in Ref.~\onlinecite{kinoshita}, the dynamics in a particular
experiment very closely resembles what we expect from an integrable
model, and it is believed that we can indeed create integrable systems
to a close approximation. This also opens up the question of how far
from integrability do we need to be in order to see the effects of
integrability breaking. We discuss this point in some detail in the
conclusions.

\section{The Lieb-Liniger model}
\label{sec:model}
Bosons in one dimensional traps interact via short range potentials,
which can be well approximated by a $\delta$-function
interaction. This model was solved in 1963 by Lieb and
Liniger~\cite{lieblin} who originally introduced it to overcome
shortcomings of other models and approaches for understanding quantum
gases and liquids, and to provide a rigorous result to test
perturbation theory against. In particular, they sought to improve
upon Girardeau's hard-core boson model~\cite{girardeau} by providing a
tunable parameter and better model a low density gas, perhaps
extensible to higher dimensions. The Schr\"odinger equation for the
model is also commonly known as the \emph{Non-linear Schr\"odinger
  equation} and has been extensively studied both classically and
quantum mechanically.

The Hamiltonian is given by
\begin{equation}
  \label{eq:llham}
  H = \int_x [\partial b^\dag(x) \partial b(x) + c \, b^\dag(x)b(x)b^\dag(x)b(x)],
\end{equation}
where $b$ is a bosonic field and $c$ is the interaction strength. The
mass has been set to 1/2.  The action for the model $\sim
\int_{x,t}[\partial_{t}-\partial_{x}^{2}]$. Time therefore has the
dimension of (length)$^{2}$. The coupling constant $c$ has the
dimensions of length.

The model is integrable and the eigenstates take the Bethe Ansatz
form,
\begin{multline}
  \label{eq:lleigen}
  \ket{\vec{\lambda}} = N(\vec{\lambda}) \int_x
  \prod_{i<j}Z_{ij}^{x}(\lambda_i-\lambda_j) \prod_j e^{i\lambda_j
    x_j}b^\dag(x_j)\ket{0},
\end{multline}
where $N(\lambda)$ is a normalization factor determined by a
particular solution, and
\begin{equation}
  Z^x_{ij}(z) = \frac{z-ic\sgn(x_i-x_j)}{z-ic}
\end{equation}
incorporates the two particle $S$-matrix, $S_{ij}=
\frac{\lambda_i-\lambda_j+ ic}{\lambda_i-\lambda_j -ic}$ occurring
when two bosons cross.  The above state satisfies
\begin{equation}
  \label{eq:llsch}
  H\ket{\vec{\lambda} }= \sum_j\lambda_j^2 \, \, \ket{\vec{\lambda}}
\end{equation}
for any value of the momenta $\vec{\lambda}$, which, depending on the
sign of $c$, may be pure real or form complex pairs. In our work, we
study the evolution dynamics of the model on the infinite line and
need not solve for explicit distributions of the $\vec{\lambda}$ that
characterize the low lying energy eigenstates, as discussed in
section~\ref{sec:opensys}.


\section{Yudson representation}
\label{sec:yudrep}

In 1985, V.~I.~Yudson presented a new approach to  time evolve
the Dicke model (a model for
superradiance in quantum optics~\cite{dicke}) considered on an infinite line~\cite{yudson1}.
The dynamics in certain cases was
extracted in closed form with much less work than previously required,
and in some cases where it was even impossible with earlier
methods. The core of the method is to bypass the laborious sum over
momenta using an appropriately chosen set of contours and integrating
over momentum variables in the complex plane. It is applicable in its
original form to models with a particular pole structure in the two
particle $S$-matrix, and a linear spectrum. We  generalize the approach to the
case of the quadratic spectrum and apply it to the study of quantum quenches.

As discussed earlier, in order to carry out the quench of a system given at $t=0$  in
a state $|\Psi_0\rangle$ one naturally proceeds by introducing a
``unity'' in terms of a complete set of eigenstates and then apply the
evolution operator--
\begin{equation}
  \label{eq:timeevol}
  \begin{split}
    \ket{\Psi_0, t}
    = e^{-iHt}\sum_{\{\lambda\}}\ket{\vec{\lambda}}\langle\vec{\lambda}|\Psi_0\rangle
    =\sum_{\{\lambda\}}e^{-i\epsilon(\vec{\lambda})t}\ket{\vec{\lambda}}\bra{\vec{\lambda}}\Psi_0\rangle
  \end{split}
\end{equation}
The Yudson representation overcomes the difficulties in carrying out this sum by
using an integral representation for the complete
basis directly in the infinite volume limit.

In the following two sections, we will discuss the
representation for the repulsive and attractive models. Each will
require a separate set of contours of integrations in order for the
representation to be valid.  We will notice that in the repulsive
case, it is sufficient to integrate over the real line.  The attractive case will require the use
of contours separated out in the imaginary direction (to be
qualified below) consistent with the fact that the spectrum consists of
``strings'' with momenta taking values as complex conjugate
pairs~\cite{lieblin}.

\subsection{Repulsive case}
\label{sec:repulsive}
We begin by discussing the repulsive case, $c>0$.  For this case, a
similar approach has been independently developed in
Ref.~\onlinecite{tracy} and has been used by
Lamacraft~\cite{lamacraft} to calculate noise correlations in the
repulsive model.  We will start with a generic initial state given by
\begin{equation}
  \ket{\Psi_{0}} = \int_{\vec{x}} \Phi_s(\vec{x}) \prod_{j} b^{\dag}(x_{j})\ket{0}.
\end{equation}
with $\Phi_s$ symmetrized. Using the symmetry of the bosonic
operators, we can rewrite this state in terms of N-boson coordinate
basis states,
\begin{equation}
  \ket{\Psi_{0}} = N!\int_{\vec{x}} \Phi_{s}(\vec x) \ket{\vec{x}}
\end{equation}
where,
\begin{equation}
  \label{eq:initstate}
  \ket{\vec{x}} = \theta(\vec{x}) \prod_j b^\dag(x_j)\ket{0}.
\end{equation}
with $\theta(\vec{x})= \theta(x_1>x_2>\cdots>x_N)$.  It suffices
therefore to show that we can express any coordinate basis state as an
integral over the Bethe Ansatz eigenstates
\begin{equation}
  \label{eq:yudrep}
  \ket{\vec x} = \theta(\vec{x})\int_\gamma \prod_j\frac{\mathrm{d} \lambda_j}{2\pi} A(\vec{\lambda},\vec{x})\ket{\vec{\lambda}}
\end{equation}
with appropriately chosen contours of integration $\{ \gamma_j \}$ and
$A(\lambda,\vec{x})$, which plays a role similar to the overlap of the
eigenstates and the initial state.

We claim that in the repulsive case equation~\eqref{eq:yudrep} is
realized with
\begin{equation}
  A(\vec{\lambda},\vec{x}) = \prod_j e^{-i\lambda_j x_j}.
\end{equation}
and the contours $\gamma_j$ running along the real axis from minus to
plus infinity. In other words eqn.~\eqref{eq:yudrep} takes the form,
\begin{multline}
  \ket{\vec x} = \theta(\vec{x})\int_{\vec y}\int \prod_{j}\frac{{\rm d}\lambda_{j}}{2\pi}\: \prod_{i<j} Z^y_{ij}(\lambda_i-\lambda_j)\\
  \times\prod_j e^{i\lambda_j(y_j-x_j)}b^\dag(y_j)\ket{0}.
\end{multline}
Equivalently, we claim that the $\lambda$ integration above produces
$\prod_{j} \delta(y_j-x_j)$.

We shall prove this in two stages. Consider first $y_N>x_N$. To carry
out the integral using the residue theorem, we have to close the
integration contour in $\lambda_N$ in the upper half plane. The poles
in $\lambda_N$ are at $\lambda_j-ic$, $j<N$. These are all below
$\gamma_N$ and so the result is zero. This implies that any non-zero
contribution comes from $y_N\leq x_N$. Let us now consider
$y_{N-1}>x_{N-1}$. The only pole above the contour is $\lambda_{N-1}^*
= \lambda_N+ic$. However, we also have, $y_{N-1}>x_{N-1}>x_{N}\geq y_N
\implies y_{N-1}>y_N$. This causes the only contributing pole to get
canceled. The integral is again zero unless $y_{N-1}\leq x_{N-1}$. We
can proceed is this fashion for the remaining variables thus showing
that the integral is non-zero only for $y_j\leq x_j$.

Now consider $y_1<x_1$. We have to close the contour for $\lambda_1$
below. There are no poles in that region, and the residue is
zero. Thus the integral is non-zero for only $y_1=x_1$. Consider
$y_2<x_2$. The only pole below, at $\lambda_2^*=\lambda_1-ic$ is
canceled as before since we have $y_2<x_2<x_1=y_1$.  Again we get that
the integral is only non-zero for $y_2=x_2$. Carrying this on, we end
up with
\begin{equation}
  \theta(\vec{x})\int_{\vec y} \prod_j \delta(y_j-x_j) b^\dag(y_j)\ket{0} = \ket{\vec x}.
\end{equation}

In order to time evolve this state, we can act on it with the unitary
time evolution operator. Since the integrals are well-defined we can
move the operator inside the integral signs to obtain,
\begin{equation}
  \ket{\vec x,t} = \theta(\vec{x})\int \prod_{j}\frac{{\rm d}\lambda_{j}}{2\pi}\:
  e^{-i\epsilon(\vec\lambda)t} A(\vec\lambda,\vec x)\ket{\vec{\lambda}}.
\end{equation}

\subsection{Attractive case}
\label{sec:attract}
We now consider an attractive interaction, $c<0$. As mentioned
earlier, this changes the spectrum of the Hamiltonian, allowing
complex, so-called \emph{string solutions}, which in this model,
correspond to many-body bound states. In fact, the ground state at
$T=0$ consists of one $N$-particle bound state.  We will see how the
Yudson integral representation takes this into account. Similar
properties are seen to emerge in~\cite{prolhac}, where the authors
obtain a propagator for the attractive Lieb-Liniger model by
analytically continuing the results obtained by Tracy and
Widom~\cite{tracy} for the repulsive model.

\begin{figure}[tb]
  \begin{center}
    \includegraphics[width=8.5cm]{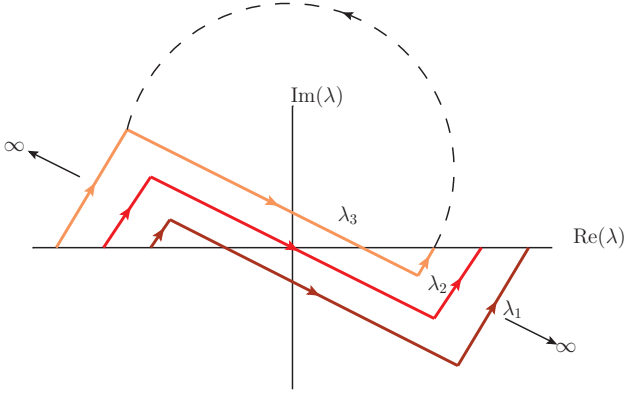}
    \caption{Contours for the $\lambda$ integration. Shown here are
      three contours, and the closing of the $N$th (here, third) contour as
      discussed in the proof.}
    \label{fig:LLcontours}
  \end{center}
\end{figure}

One will immediately notice from the eigenstates that the pole
structure of the $S$-matrix is altered. This change prevents the proof
of the previous section from working. In particular, the poles in the
variable $\lambda_{N}$ are at $\lambda_{j}+i\abs{c}$ for $j<N$, and
the residue of the contour closed in the upper half plane is not zero
any more. We need choose a contour to avoid this pole. This can be
achieved by separating the contours in the imaginary direction such
that adjacent ${\rm Im}[\lambda_{j}-\lambda_{j-1}] > \abs{c}$. At
first sight, this seems to pose a problem as the quadratic term in
exponent diverges at large positive $\lambda$ and positive imaginary
part.  There are two ways around this. We can tilt the contours as
shown in Fig.~\ref{fig:LLcontours} so that they lie in the convergent
region of the Gaussian integral. The pieces towards the end, that join
the real axes though essential for the proof to work at $t=0$, where
we evaluate the integrals using the residue theorem, do not contribute
at finite time as the integrand vanishes on them as they are taken to
infinity.
Another more natural means of doing this is to use the finite spatial
support of the initial state. The overlaps of the eigenstates with the
initial state effectively restricts the support for the $\lambda$
integrals, making them convergent.

The proof of equation~\eqref{eq:yudrep} now proceeds as in the
repulsive case. We start by assuming that $y_N>x_N$ requiring us to
close the contour in $\lambda_N$ in the upper half plane. This
encloses no poles due to the choice of contours and the integral is
zero unless $y_N\leq x_N$. Now assume $y_{N-1}>x_{N-1}$.  Closing the
contour above encloses one pole at $\lambda_{N-1}^*=
\lambda_N-i\abs{c}$, however since $y_{N-1}>x_{n-1}>x_N\geq y_N$, this
pole is canceled by the numerator and again we have $y_{N-1}\leq
x_{N-1}$. We proceed in this fashion and then backwards to show that
the integral is non-zero only when all the poles cancel, giving us
$\prod_j \delta(y_j-x_j)$, as required.

\section{Two particle dynamics}

\label{sec:2pquench}

We begin with a detailed discussion of the quench dynamics of two
bosons. As we saw, it is convenient to express any initial state in
terms of an ordered coordinate basis, $ \ket{\vec{x}} =
\theta(x_1>x_2>\cdots>x_N)\prod_j b^\dag(x_j)\ket{0}$.  At finite
time, the wave function of bosons initially localized at $x_1$ and
$x_2$ and subsequently evolved by a repulsive Lieb-Liniger Hamiltonian
is given by,
\begin{equation}
  \label{eq:rep2p}
  \begin{split}
    &\ket{\vec{x},t}_2 = e^{-iHt}\theta(x_1-x_2) b^{\dagger}(x_1) b^{\dagger}(x_2) \ket{0} \\
    &=\int_{y,\lambda} Z^y_{12}(\lambda_1-\lambda_2)  e^{-i\lambda_1^2t-i\lambda_2^2t+i\lambda_1(y_1-x_1)+i\lambda_2(y_2-x_2)}\\
    &\qquad\qquad \times b^\dag(y_1)b^\dag(y_2)\ket{0}\\
    &= \int_y \frac{e^{i\frac{(y_1-x_1)^2}{4t}+i\frac{(y_2-x_2)^2}{4t}}}{4\pi i t}\\
    &\times\Big[1- c\sqrt{\pi i
      t}\theta(y_2-y_1)e^{\frac{i}{8t}\alpha^2}
    \erfc\left(\frac{i-1}{4}\frac{i\alpha}{\sqrt{t}}
    \right)\Big]\\
    &\times b^\dag(y_1)b^\dag(y_2)\ket{0}.
  \end{split}
  \raisetag{80pt}
\end{equation}
where $\alpha = 2ct-i(y_1-x_1)-i(y_2-x_2)$.  The above expression
retains the Bethe form of wave functions defined in different
configuration sectors. The only scales in the problem are the
interaction strength $c$ and the scale from the initial condition, the
separation between the particles at $t=0$.

In order to get physically meaningful results we need to start from a
physical initial state. We choose first the state
$\ket{\Psi(0)_{\text{latt}}}$ where bosons are trapped in a periodic
trap forming initially a lattice-like state (see
fig.~\ref{fig:initstate}a),
\begin{equation}
  \label{eq:harmtrap}
  \ket{\Psi(0)_{\text{latt}}} = \prod_j   \left[ \frac{1}{(\pi\sigma^2)^{\frac14} } \int_{\vec{x}}e^{-\frac{(x_j+(j-1)a)^2}{2\sigma^2}}   b^\dag(x_{j})\right]\ket{0}.
\end{equation}
If we assume that the wave functions of neighboring bosons do not
overlap significantly, i.e., $ e^{-\frac{a^2}{\sigma^2}} \ll 1$, then
the ordering of the initial particles needed for the Yudson
representation is induced by the non-overlapping support and it
becomes possible to carry out the integral analytically.

\begin{figure}[htb]
  \centering \includegraphics[width=8.7cm]{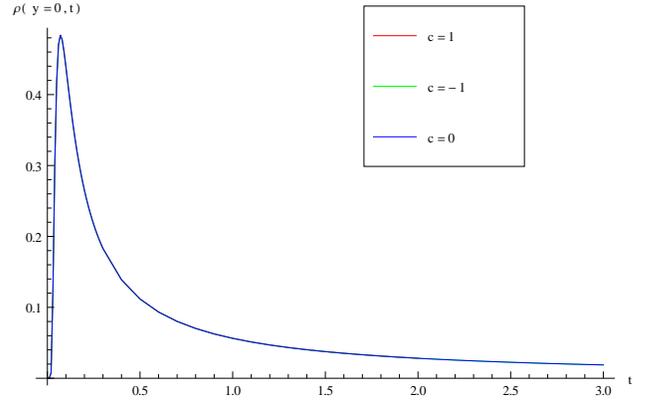}
  \caption{(Color online) $\langle\rho(x=0,t)\rangle$ vs. $t$, after
    the quench from $\ket{\Psi_{\text{latt}}}$. $\sigma/a\sim
    0.1$. The curves appear indistinguishable (i.e. lie on top of each other) since the particles
    start out with non significant overlap. The interaction effects
    would show up only when they have propagated long enough to have spread
    sufficiently to reach a significant overlap, at which time the
    density is too low.}
  \label{fig:twores}
\end{figure}

We now calculate the evolution of some observable in the state
$\ket{\Psi(0)_{\text{latt}}}$. Consider first the evolution of the
density $\rho(x)=b^{\dag}(x)b(x)$ at $x=0$. Fig.~\ref{fig:twores}
shows $\bra{\Psi_{\text{latt}} ,t} \rho(0)\ket{\Psi_{\text{latt}},t}$
for repulsive, attractive and non-interacting bosons. No difference is
discernible between the three cases. The reason is obvious: the local
interaction is operative only when the wave functions of the particles
overlap. As we have taken $\sigma \ll a$ this will occur only after a
long time when the wave-function is spread out and overlap is
negligible.

\begin{figure}[htb]
  \centering \includegraphics[width=8.5cm]{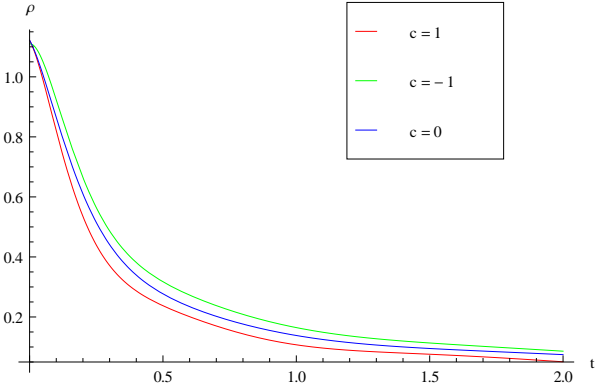}
  \caption{(Color online) $\langle\rho(x=0,t)\rangle$ vs. $t$, after
    the quench from $\ket{\Psi_{\text{cond}}}$. $\sigma \sim 0.5,\:
    a=0$. As the bosons overlap interaction effects show up
    immediately. Lower line: $c=1$, Upper line: $c=-1$, Middle line: $c=0$. }
  \label{fig:tworescond}
\end{figure}

Consider now an initial state where we set the separation $a$ to zero,
starting with maximal initial overlap between the bosons $|
\Psi(0)_{\text{cond}}\rangle$. We refer to this state as a condensate
(in position-space). Fig.~\ref{fig:tworescond} shows the density
evolution for attractive, repulsive and no interaction.  The decay of
the density is slower for attractive model than the for the
non-interacting which in turn is slower than for the repulsive model -
indeed, unlike before, the interaction is operative from the
beginning.
Still, the density does not show much difference between repulsive and
attractive interactions in this case. However, a drastic difference
will appear when we study the noise correlations $\langle \Psi_0,t
|\rho(x_1)\rho(x_2)|\Psi_0,t \rangle$, as will be shown below.

A comment about the attractive case is in order here.  Recall that the
contours of integration are separated in the imaginary direction.  In
order to carry out the integration over $\lambda$, we shift the
contour for $\lambda_{2}$ to the real axis, and add the residue of the
pole at $\lambda_{2}=\lambda_{1}+i\abs{c}$.  The two particle finite
time state can be written as
\begin{equation}
  \begin{split}\label{eq:att2pint}
    &\ket{\vec{x},t}_{2}
    =\int_y\int_{\gamma_c}\prod_{i<j=1,2}Z^y_{ij}(\lambda_i-\lambda_j)\\
    &\qquad\qquad \times \prod_{j=1,2}
    e^{-i\lambda_{j}^{2}t+i\lambda_j (y_{j}-x_j)}b^\dagger(y_j)\ket{0}\\
    &=\int_y\bigg[\int_{\gamma_r}\prod_{i<j,1}^2Z^y_{ij}(\lambda_i-\lambda_j)\prod_{j,1}^2
    e^{-i\lambda_{j}^{2}t+i\lambda_j(y_{j} -x_j)}b^\dagger(y_j)\ket{0}\\
    &\qquad + \theta(y_{2}-y_{1})I(\lambda_2=\lambda_1+i\abs{c},t)
    b^\dagger(y_1)b^\dagger(y_2)|0\rangle\bigg]
  \end{split}
  \raisetag{18pt}
\end{equation}
$\gamma_c$ refers to contours that are separated in imaginary
direction, $\gamma_r$ refers to all $\lambda$ integrated along real
axis. $I(\lambda_2=\lambda_1+i \abs{c},t)$ is the residue obtained
by shifting the $\lambda_{2}$ contour to the real axis from the pole
at $\lambda_2$ at $\lambda_1+i\abs{c}$. This second term corresponds
to a two-particle bound state. It is given by
\begin{equation}
  \begin{split}\label{eq:att2pbd}
    &I(\lambda_2=\lambda_1+i \abs{c},t) =\\
    &-2c\int_y\int_{\lambda_1}e^{i\lambda_1(y_1-x_1)+i(\lambda_{1}+i\abs{c})(y_2-x_2)-i\lambda_{1}^{2}t - i(\lambda_{1}+i\abs{c})^{2}t}\\
    &=-2c\int_y\int_{\lambda_1}e^{i\lambda_1(y_1-x_1+y_2-x_2)-\frac{\abs{c}}{2}(y_2-y_1)-\frac{\abs{c}}{2}(x_1-x_2)}\\
    & \qquad\qquad \times e^{-i(\lambda_{1}-i\abs{c}/2)^{2}t - i(\lambda_{1}+i\abs{c}/2)^{2}t}\\
    &=-c\int_y\int_{\lambda_1}e^{i\lambda_1(y_1-x_1+y_2-x_2)-\frac{\abs{c}}{2}(x_1-x_2) -\frac{\abs{c}}{2}|y_2-y_1|}\\
    & \qquad\qquad\times e^{-2i\lambda_{1}^{2}t + i\frac{\abs{c}^{2}}{2}t}.
  \end{split}
\end{equation}
This contribution corresponds to the particles propagating as a bound
state, $e^{-\frac{\abs{c}}{2}|y_2-y_1|}$, with kinetic energy $E_k=
2\lambda_{1}^{2}$, and binding energy $E_b= -c^{2}/2$. The rest of the
expression,
$e^{i\lambda_1(y_1-x_1+y_2-x_2)-\frac{\abs{c}}{2}(x_1-x_2)}$, yields
the overlap of the bound state with the initial state
$|\vec{x}\rangle$. Note that the overlap decays exponentially as the
distance $\abs{x_1-x_2}$ between the initial positions is increased.

Such bound states appear for any number of particles involved. For
instance, for three particles, the Yudson representation with complex
$\lambda$s automatically produces multiple bound-states coming from
the poles, i.e. $ I(\lambda_2=\lambda_1+i\abs{c})$,
$I(\lambda_3=\lambda_1+i\abs{c})$, etc. They give rise to two and
three particle bound states, the latter being of the form
$e^{-\frac{|c|}{2}(|y_1-y_2|+|y_1-y_3|+|y_2-y_3|)}$ with binding
energy $E_b=-2c^2$.  It is important to remark that these bound states
were not put in by hand, but arise straightforwardly from the
contour representation. The binding energy of an $N$-particle bound
state thus appearing in the time evolution is $E_b= -c^2 N(N^2-1)/12$
as expected from the spectrum of the Hamiltonian~\cite{yang}.

Finally, we combine the  bound state contribution discussed above
and the ``real axis'' term,
$\int_y\int_{\gamma_r}\prod_{i<j,1}^2Z^y_{ij}(\lambda_i-\lambda_j)\prod_{j,1}^2
e^{-i\lambda_{j}^{2}t+i\lambda_j(y_{j} -x_j)}b^\dagger(y_j)\ket{0}$,
which corresponds to the non-bound propagating states in
\eqref{eq:att2pint}. The resulting wave function is,
\begin{equation}
  \label{eq:att2p}
  \begin{split}
    &\ket{\vec{x},t}_2
    = \int_y \frac{e^{i\frac{(y_1-x_1)^2}{4t}+i\frac{(y_2-x_2)^2}{4t}}}{4\pi i t}\\
    &\left[1+ \abs{c}\sqrt{\pi i
        t}\theta(y_2-y_1)e^{\frac{i}{8t}\tilde\alpha^2}\erfc\left(\frac{i-1}{4}\frac{i\tilde\alpha}{\sqrt{t}}
      \right)\right]\\
    &\qquad\qquad\qquad\times b^\dagger(y_1)b^\dagger(y_2)|0\rangle
  \end{split}
\end{equation}
where $\tilde\alpha = -2\abs{c}t-i(y_1-x_1)-i(y_2-x_2)$. Surprisingly,
the wave function maintains its form and we only need to replace $c
\to -c$.  This simple result is not valid for more than two particles.

\begin{figure}[hbt]
  \centering \includegraphics[width=7.5cm]{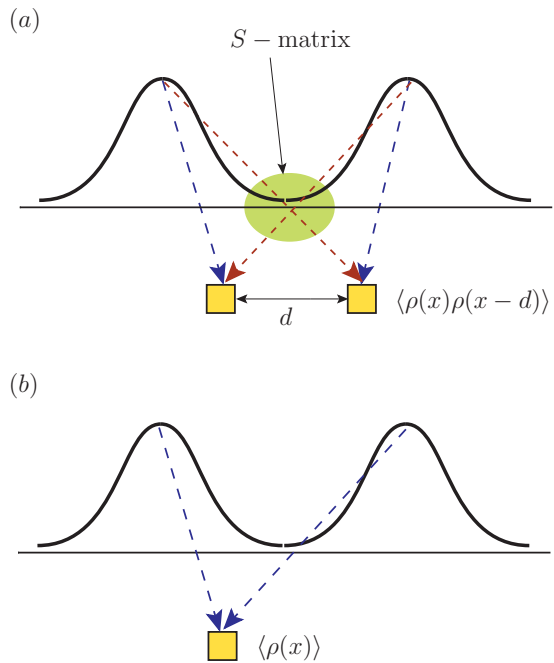}
  \caption{$(a)$  The Hanbury-Brown Twiss effect, where
    two detectors are used to measure the interference of the direct (big dashes)
    and the crossed waves (small dashes). The $S$-matrix enters explicitly. $(b)$ The
    density measurement is not directly sensitive to the
    $S$-matrix. The thick black line shows the wave-function
    amplitude, the dotted lines show time propagation.}
  \label{fig:hbt}
\end{figure}

We now compute the noise correlation in the evolving state. We expect
the interaction to have a a significant effect as the geometry of the
set up measures the interference of ``direct'' and ``crossed''
measurements as shown in fig.~\ref{fig:hbt}a.
In contrast, the density measurements do not see the $S$-matrix, as
shown in fig.~\ref{fig:hbt}$b$.

\begin{figure}[htb]
  \centering \includegraphics[width=7cm]{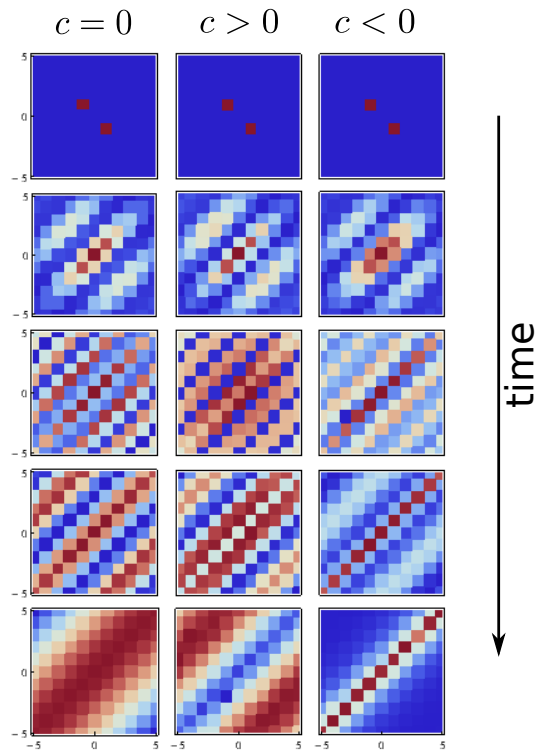}
  \caption{(Color online) Time evolution of density-density
    correlation matrix ($\langle \rho(x)\rho(y)\rangle$) for the
    $\ket{\Psi_{\text{latt}}}$ initial state.  Blue is zero and red is
    positive. The repulsive model shows anti-bunching, i.e.,
    fermionization at long times, while the attractive model shows
    bunching.}
  \label{fig:corr2platt}
\end{figure}

\begin{figure}[htb]
  \centering \includegraphics[width=7cm]{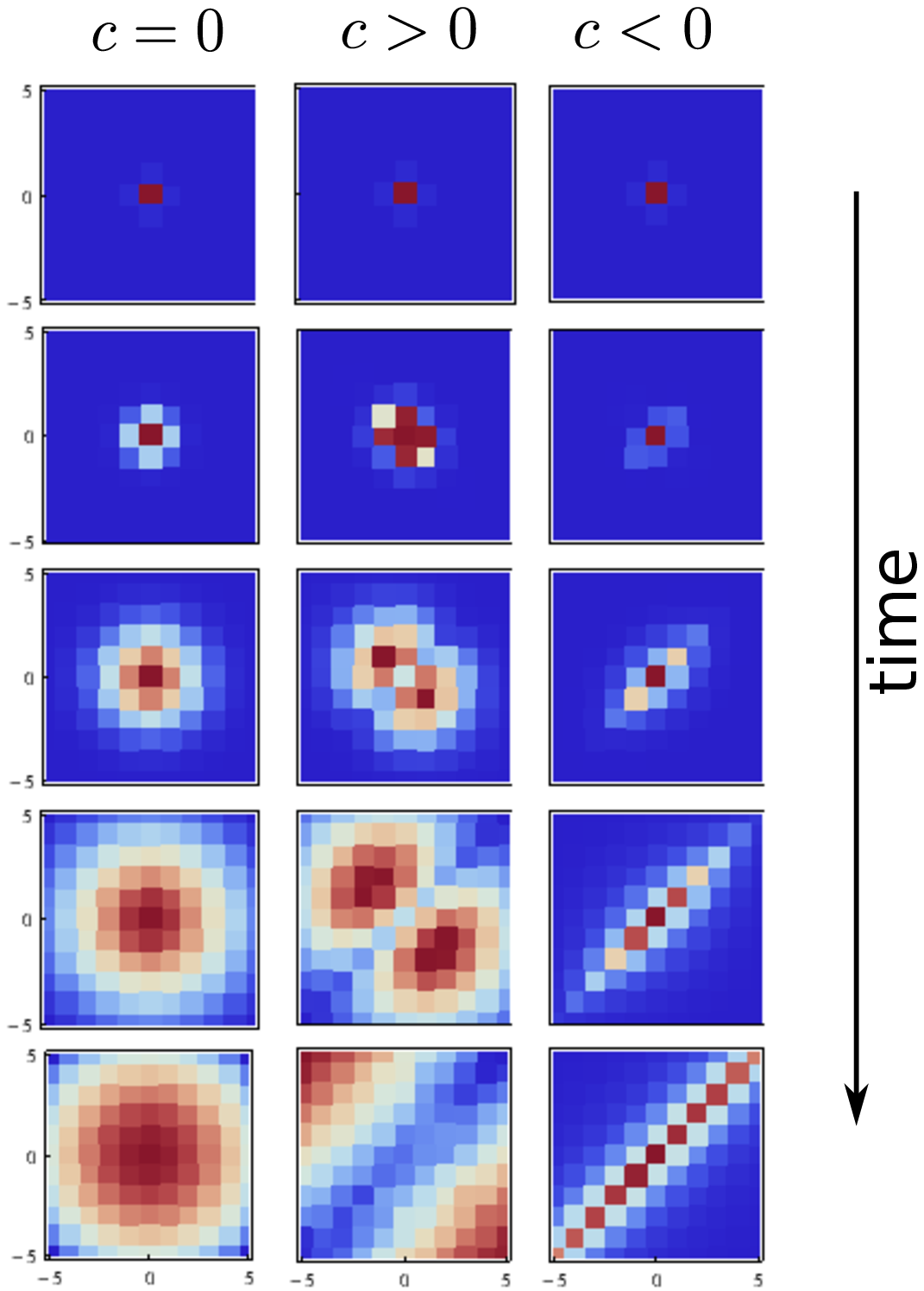}
  \caption{(Color online) Time evolution of density-density
    correlation matrix ($\langle \rho(x)\rho(y)\rangle$) for the
    $\ket{\Psi_{\text{cond}}}$ initial state.  Blue is zero and red is
    positive. The repulsive model shows anti-bunching, i.e.,
    fermionization at long times, while the attractive model shows
    bunching.}
  \label{fig:corr2pcond}
\end{figure}

This is the famous Hanbury-Brown Twiss experiment~\cite{hbt} where for
free bosons or fermions, the crossing produces a phase of $\pm 1$ and
causes destructive or constructive interference. In our case the set
up is generalized to multiple time dependent sources with the phase
given by the two particle $S$-matrix capturing the interactions
between the particles.  In Fig.~\ref{fig:corr2platt} we present the
two point correlation matrix $\langle \rho(x_1)\rho(x_2)\rangle$ for
the repulsive gas, attractive gas, and the non-interacting gas, shown
at different times, starting with the lattice initial
state. Figure~\ref{fig:corr2pcond} shows the same for the condensate
initial state.
In both these, we note that the repulsive gas develops fermionic
correlations (i.e., strong anti-bunching), and the attractive gas
retains bosonic correlations at long time, showing strong bunching.

It is interesting to compare this result with the time evolution after
a quench on the lattice by the Bose-Hubbard model, the lattice
counterpart of the Lieb-Liniger model, as we shall see in
Appendix~\ref{sec:bhtebd}. The results for continuum model differs
strongly from those of the lattice model.

We expect the results to be qualitatively similar for higher particle
number.  In order to go beyond two particles however, the integrations
cannot be carried out exactly. However, we can extract the asymptotic
behavior of the wavefunction analytically, as we show below.

\section{Multiparticle dynamics at long times}
\label{sec:asymp}

In this section, we derive an expression for the multiparticle
wavefunction evolution at long times.  The number of particles $N$ is
kept fixed in the limiting process, hence, as discussed in the
introduction we are in the low density limit where interactions are
expected to be dominant.  The other regime where $N$ is sent to
infinity first will be discussed in a separate report.  We first deal
with the repulsive model, for which  no bound states exist and
the momentum integrations can be carried out over the real line, and
then proceed to the attractive model. In a separate sub-section, we
examine the effect of starting with a condensate-like initial state.

\subsection{Repulsive interactions - Asymptotics}
\label{sec:repasymp}
From \eqref{eq:yudrep} we can see by scaling
$\lambda\to\lambda\sqrt{t}$, we get
\begin{equation}
  Z^y_{ij}(\lambda_i-\lambda_j) \to \sgn(y_i-y_j)+ O\left(\frac{1}{\sqrt{t}}\right),
\end{equation}
yielding to leading order,
\begin{equation*}
  \begin{split}
    \label{eq:claimt}
    &|\Psi_0,t\rangle \to
    \int_x \int_{y}\int_{\lambda}\,\theta(\vec{x})\Psi_0(\vec{x})\\
    &\times \prod_j
    \frac{1}{\sqrt{t}}e^{-i\lambda_j^2+i\lambda_j(y_j-x_j)/\sqrt{t}}
    \prod_{i<j} \sgn(y_i-y_j) b^\dag(y_j)\ket{0}\\
    &= \int_{x,y,\lambda,k} \,\theta(\vec{x})\Psi_0(\vec{x}) \prod_j
    e^{-i\lambda_j^2t+i\lambda_j(y_j-x_j)}
    e^{-ik_{j}y_{j}}c^{\dag}_{k_{j}}\ket{0}\\
    &=\int_{x,k} \,\theta(\vec{x})\Psi_0(\vec{x}) \prod_j e^{-ik_j^2t-ik_{j} x_j}c^{\dag}_{k_{j}}\ket{0}\\
    &= e^{-iH^f_0
      t}\int_{x}\mathcal{A}_x\theta(\vec{x})\Psi_0(\vec{x}) \prod_j
    c^\dag(x_j)\ket{0},
  \end{split}
\end{equation*}
$c^{\dagger}(y)$ being fermionic creation operators replacing the
``fermionized'' hardcore bosonic operators, $\prod_j
c^{\dagger}(y_j)=\prod_{i<j} \sgn(y_i-y_j) b^\dag(y_j)$. We denote
$H^f_0=\int_x \partial c^{\dagger}(x)\partial c(x)$ the free fermionic
Hamiltonian and $\mathcal{A}_y$ is an anti-symmetrizer acting on the
$y$ variables.  Thus, the repulsive Bose gas, for \emph{any} value of
$c>0$, is governed in the long time by the $c=\infty$ hard core boson
limit (or its fermionic equivalent)~\cite{jukic,girardeau}, and the
system equilibrates with an asymptotic momentum distribution,
$n_k=\langle \tilde{\Psi}_0|c^{\dagger}_kc_k|\tilde{\Psi}_0\rangle$,
determined by the antisymmetric wavefunction
$\tilde{\Psi}_0(\vec{y})=\mathcal{A}_y \theta(\vec{y})\Psi_0(\vec{y})$
and the total energy, $E_{\Psi_0}=\langle\Psi_0\rvert H
\lvert\Psi_0\rangle$.

We will now derive the corrections to the infinite time limit. At
large time, we use the stationary phase approximation to carry out the
$\lambda$ integrations.  The phase oscillations come primarily from
the exponent $e^{-i\lambda_{j}^{2}t +i\lambda_{j}(y_{j}-x_{j})}$.  At
large $t$ (i.e., $t\gg\frac{1}{c^{2}}$), the oscillations are rapid,
and the stationary point is obtained by solving
\begin{equation}
  \frac{\rmd}{\rmd \lambda_{j}} [-i\lambda_{j}^{2}t + i\lambda_{j}(y_{j}-x_{j})]= 0.
\end{equation}
Note that typically one would ignore the second term above since it
doesn't oscillate faster with increasing $t$, but here we cannot since
the integral over $y$ produces a non-zero contribution for $y\sim t$
at large time.  Doing the Gaussian integral around this point (and
fixing the $S$-matrix prefactor to its stationary value), we obtain
for the repulsive case,
\begin{multline}
  \label{eq:inftime}
  \ket{\vec{x},t}  \to  \int_y \prod_{i<j} Z^y_{ij}\left(\frac{y_i-y_j-x_i+x_j}{2t}\right)\\
  \times\prod_j \frac{1}{\sqrt{4\pi
      it}}e^{-i\frac{(y_j-x_j)^2}{4t}+i\frac{(y_j-x_j)^2}{2t}}
  b^\dag(y_j)\ket{0}.
\end{multline}
In the above expression, the wavefunction has support mainly from
regions where $y_j/t$ is of order one.  In an experimental setup, one
typically starts with a local finite density gas, i.e., a finite
number of particles localized over a finite length. With this
condition, at long time, we can neglect $x_{j}/t$ in comparison with
$y_{j}/t$, giving
\begin{equation}
  \label{eq:inftime2}
  \ket{\vec{x},t} \to  \int_y \prod_{i<j} Z_{ij}(\xi_i-\xi_j)
  \prod_j \frac{1}{\sqrt{4\pi it}}e^{it\xi_j^2-i\xi_j x_j} b^\dag(y_j)\ket{0}
\end{equation}
where $\xi=\frac{y}{2t}$.

We turn now to calculated the asymptotic evolution of some
observables.  To compute the expectation value of the density we start
from the coordinate basis states, $\bra{\vec{x'},t
}\rho(z)\ket{\vec{x},t}$ which we then integrate with the chosen
initial state,
\begin{multline}
  \bra{\vec{x'},t }\rho(z)\ket{\vec{x},t} =
  \sum_{\{P\}}\int_y\left(\sum_j\delta(y_j-z)\right) \\
  \times\prod_{i<j} Z_{ij}(\xi_i-\xi_j)Z^*_{P_i
    P_j}(\xi_{P_i}-\xi_{P_j}) \prod_j \frac{1}{4\pi t}e^{-i(\xi_j
    x_j-\xi_{P_j}x'_j)}
\end{multline}

Note that the above product of $S$-matrices is actually independent of
the ordering of the $y$. First, only those terms appear in the product
for which the permutation $P$ has an inversion. For example, say for
three particles, if $P=312$, then the inversions are 13 and 23. It is
only these terms which give a non-trivial $S$-matrix contribution.
For the non-inverted terms, here 12, we get
\begin{equation}
  \frac{\xi_1-\xi_2-ic\sgn(y_1-y_2)}{\xi_1-\xi_2-ic} \frac{\xi_1-\xi_2+ic\sgn(y_1-y_2)}{\xi_1-\xi_2+ic}
\end{equation}
which is always unity irrespective of the ordering of $y_1,y_2$. For a
term with an inversion, say 23, we get,
\begin{equation}
  \frac{\xi_2-\xi_3-ic\sgn(y_2-y_3)}{\xi_2-\xi_3-ic} \frac{\xi_3-\xi_2+ic\sgn(y_3-y_2)}{\xi_3-\xi_2+ic}
\end{equation}
which is always equal to
\begin{equation}
  \frac{\xi_2-\xi_3+ic}{\xi_2-\xi_3-ic}\equiv S(\xi_2-\xi_3)
\end{equation}
irrespective of the sign of $y_2-y_3$. This allows us to carry out the
integration over the $y_j$.

\subsubsection{Lattice initial state}

In order to calculate physical observables, we have to choose initial
states.  We choose two different initial states for the problem, one
with $N$ particles distributed with uniform density in a series of
harmonic traps given by,
\begin{equation}
  \label{eq:init1}
  \ket{\Psi_{\text{latt}}} = \int_x \prod_{j=1}^N \frac{1}{(\pi \sigma^2)^{\frac14}}e^{-\frac{(x_j+(j-1)a)^2}{2\sigma^2}}b^\dag(x_j)\ket{0},
\end{equation}
such that the overlap between the wave functions of two neighboring
particles is negligible. In this particular case, the ordering of the
particles is induced by the limited non-overlapping support of the
wave function.

In the lattice-like state, the initial wave function starts out with
the neighboring particles having negligible overlap. At small time (as
seen from \eqref{eq:rep2p}), the particle repel each other, but they
never cross due to the repulsive interaction. So at large time, the
interaction does not play a role since the wave functions are
sufficiently non-overlapping. It is only the $P=1$ contribution then
that survives, and we get for the density
\begin{equation}
  \bra{\vec{x}',t}\rho(z)\ket{\vec{x},t} =
  \sum_j \prod_{k\ne j}\frac{\delta(x_k-x_k')e^{-i \frac{z}{2t}(x_j-x_j')}}{4\pi t}.
\end{equation}
We need to integrate the position basis vectors $\ket{\vec{x}}$ over
some initial condition. We do this here for the lattice
state~\eqref{eq:init1} This gives
\begin{equation}
  \bra{\Psi_{\text{latt}},t} \rho(z) \ket{\Psi_{\text{latt}},t}=\rho_{\text{latt}}(\xi_z) = \frac{N\sigma}{2\sqrt{\pi}t}e^{-\frac{\xi_z^2}{\sigma^2}}
\end{equation}

Mathematically, any $S$-matrix factor that appears will necessarily
have zero contribution from the pole - this is easy to see from the
pole structure, and the ordering of the coordinates.  In order to get
a non-zero result, we need to fix at least two integration variables
(i.e., the $y_{j}$). Thus the first non-trivial contribution comes
from the two-point correlation function.

We now proceed to calculate the evolution of the noise, i.e., the two
body correlation function $\rho_2(z,z'; t)_{\text{latt}} =
\bra{\Psi_{\text{latt}},t}\rho(z)\rho(z') \ket{\Psi_{\text{latt}},t}$.
 The contributions can
be grouped in terms of number of crossings, which corresponds to
a grouping in terms of the coefficient $e^{-ca}$~\cite{lamacraft}. The leading order
term can be explicitly evaluated and we show below which terms
contribute. In general we have
\begin{multline}
  \rho_{2\text{ latt}}(z,z';t) = \sum_{\{P\}}\int_y\left(\sum_{j,K}\delta(y_j-z)\delta(y_k-z')\right)\\
  \times \prod_{i<j,(ij)\in P} S(\xi_i-\xi_j) \prod_j \frac{1}{4\pi
    t}e^{-i(\xi_j x_j-\xi_{P_j}x'_j)}.
\end{multline}
The above shorthand in the $S$-matrix product means that only the
$(ij)$ that belong to the inversions in $P$ are included. We will now
determine which terms contribute in this sum.  First note that for
integration over a particular $\xi_j$, the residue depends on the sign
of $x_j-x_{P_j}$. Let us consider a specific example. Consider the
three particle case with the term $P=321$. All three $S$-matrix
factors appear in this term. $\xi_3$ has a pole at $\xi_1-ic$ and
$\xi_2-ic$. Thus integrating over $y_3$ will give a non-zero residue
only if $x_3>x_1'$ which is however not satisfied by the initial
conditions we choose. So, everything is zero, unless we do not
integrate over $y_3$, implying it has to be one of the measured
variables. Similarly for $\xi_1$, the poles are at $\xi_2+ic$ and
$\xi_3+ic$. In order to get a non-zero residue we need $x_1<x_3'$
which again is not satisfied by the initial conditions. We get a
non-zero result if we pin $\xi_1$. As for $\xi_2$ it has poles both
above and below the real line and so this always gives a non-zero
contribution irrespective of the sign of $x_2-x_2'$.

One can see that this argument can be extended to the case with more
particles. Depending on what coordinates we are measuring at, we'll
get a specific contribution from the sum over permutations. The next
simplification comes from not allowing any crossings among the
unmeasured coordinates. It can be shown that allowing for these gives
us a higher order contribution in $e^{-ca}$. In other words, the
leading order contribution comes from terms such as
$P=21,32,321,4231,5342,52341,\ldots$. The only exchanges are on the
ends.  A general term will therefore look like (for $l<k$),
\begin{equation}
  \begin{split}
    & \int_y\delta(y_l-z)\delta(y_k-z')
    \prod_{j\ne\{l,k\}} S(\xi_i-\xi_j)S(\xi_j-\xi_k)\\
    & \times \frac{e^{-i\xi_j( x_j-x_j')}}{4\pi t}
    \frac{\xi_l-\xi_k+ic}{\xi_l-\xi_k-ic}\frac{e^{-i\xi_l(x_l-x_k')-i\xi_k(x_k-x_l')}}{(4\pi
      t)^2}
  \end{split}
\end{equation}
We have to sum over $l,k$, which will automatically sum over the
number of intermediate $j$'s appearing. We'll integrate the above
general term, since the $y_j$ integrals factor anyway. This gives
\begin{multline}
  \prod_{i\ne k,l,j}\delta(x_i-x'_i)S(\xi_l-\xi_k)\frac{e^{-i\xi_l(x_l-x'_k)-i\xi_k(x_k-x'_l)}}{(4\pi t)^2} \\
  \times\prod_j \bigg[\delta(x_j-x'_j)-S(\xi_l-\xi_k-ic)\\
  \times\bigg\{\theta(x_j>x'_j)e^{-i(\xi_l-ic)(x_j-x'_j)}+\\
  + \theta(x_j<x'_j)e^{-i(\xi_k+ic)(x_j-x'_j)} \bigg\}\bigg]
\end{multline}
We can sum the different contributions now. Note that the number of
terms appearing the product over $j$ is given by $k-l-1$. So for a
given $l$, we have to sum over all the $k$. Using a shorthand
notation, the sum can be written as (note that it is understood that
$y_l$ and $y_k$ are integrated over using the delta functions.  We
retain the indices to keep track of the terms. We actually have
$\xi_l=2zt$ and $\xi_k=2z't$.
\begin{equation}
  \sum_{l,k}\prod_{i<l,i>k}\delta_i f_{lk} \prod_{j=l+1}^{k-1} g_{jlk}
\end{equation}
In order to proceed with the summation, we have to integrate over the
$x$. We use the initial conditions described by \eqref{eq:init1},
i.e., the lattice-like state. We'll do it term by term.
\begin{equation}
  \int_{x_i,x'_i} \delta(x_i-x'_i) \frac{e^{-\frac{(x_i+(i-1)a)^2}{2\sigma^2}-\frac{(x'_i+(i-1)a)^2}{2\sigma^2} }}{\sqrt{\pi\sigma^2}}
  = 1
\end{equation}
\begin{multline}
  \int_{x_l,x'_k,x_k,x'_l} f_{lk} \frac{e^{-\sum_{l,k,l',k'}\frac{(x_l+(l-1)a)^2}{2\sigma^2}}}{\pi\sigma^2}\\
  = \frac{\sigma^2}{4\pi t^2}S(\xi_l-\xi_k)
  e^{-(\xi_l^2+\xi_k^2)\sigma^2}e^{ia(k-l)(\xi_k-\xi_l)}
\end{multline}
\begin{multline}
  \int_{x_j,x'_j} g_{jlk}
  \frac{e^{-\frac{(x_j+(j-1)a)^2}{2\sigma^2}-\frac{(x'_j+(j-1)a)^2}{2\sigma^2}
    }}{\pi\sigma^2}
  =  1-2c\sqrt{\pi}\sigma \\
  \times S(\xi_l-\xi_k-ic)\Big[e^{(c+i\xi_l)^2\sigma^2}\erfc\{(c+i\xi_l)\sigma\}+\\
  + e^{(c-i\xi_k)^2\sigma^2}\erfc\{(c-i\xi_k)\sigma\}\Big]
\end{multline}
Note that last expression has no $j$ dependence. So, the sum over the
product over $j$ is just a geometric series. Recall that $\xi_l$ and
$\xi_k$ are fixed at $z$ and $z'$ respectively. This series can be
summed. Note that the number of terms in the product is equal to the
$k-l$, and summing over them is effectively summing over $k$. The
previous term therefore needs to be taken into account.  Writing
$\xi_l=\xi_z$ and $\xi_k=\xi_{z'}$, we can write the sum as
\begin{equation}
  \frac{\sigma^2}{4\pi t^2}S(\xi_z-\xi_{z'})e^{-(\xi_x^2+\xi_{z'}^2)\sigma^2}
  \sum_{l<k} e^{ia(k-l)(\xi_z-\xi_{z'})} g_{zz'}^{k-l-1}
\end{equation}
where
\begin{multline}
  g_{zz'} =
  1-2c\sqrt{\pi}\sigma S(\xi_z-\xi_{z'}-ic)\\
  \times\Big[e^{(c+i\xi_z)^2\sigma^2}\erfc\{(c+i\xi_z)\sigma\}+\\
  +e^{(c-i\xi_{z'})^2\sigma^2}\erfc\{(c-i\xi_{z'})\sigma\}\Big]
\end{multline}

\begin{figure}[h]
  \centering \includegraphics[width=8cm]{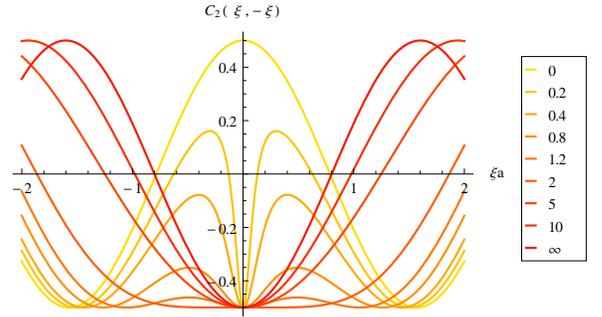}
  \caption{(Color online) Normalized noise correlation function
    $C_2(\xi,-\xi)$.  Fermionic correlations develop on a time scale
    $\tau\sim c^{-2}$, so that for any $c$ we get a sharp fermionic
    peak near $\xi=0$, i.e., at large time.  The key shows values of
    $ca$ (from Ref.~\onlinecite{deeprl}).}
  \label{fig:largec}
\end{figure}

Finally, we have to account for the $k>l$ case which is equivalent to
setting $\xi_l=\xi_{z'}$ and $\xi_l=\xi_z$. Doing this is further
equivalent to adding the complex conjugate. We also have to take into
account the term with no permutations. So, finally we have,
\begin{multline}
  \rho_{2 \text{ latt}}(z,z') = \frac{N^2\sigma^2}{4\pi
    t^2}e^{-(\xi_x^2+\xi_{z'}^2)\sigma^2}\bigg[1+
  \frac{2}{N^2}\mathrm{Re}S(\xi_{z}-\xi_{z'}) \\
  \times e^{ia(z-z')}
  \frac{N(1-e^{ia(z-z')}g)+e^{iaN(z-z')}g_{zz'}^N-1}{[1-g_{zz'}e^{ia(z-z')}]^2}\bigg]
\end{multline}
To compare with the Hanbury-Brown Twiss result, we calculate the
normalized spatial noise correlations, given by $C_2(z,z')\equiv
\frac{\rho_2(z,z')}{\rho(z)\rho(z')} - 1\equiv C_2(z,z')$.  In the
non-interacting case, i.e., $c=0$, $S(\xi)=1$ and $g_{zz'}=0$ and we
recover the HBT result for $N=2$,
\begin{equation}
  C_2^0(\xi_z,\xi_{z'}) = \frac12 \cos(a(\xi_z-\xi_z'))
\end{equation}
One can also check that the limit of $c\to\infty$ gives the expected
answer for free fermions, namely,
\begin{equation}
  C_2^\infty(\xi_z,\xi_{z'}) = -\frac12 \cos(a(\xi_z-\xi_z'))
\end{equation}
At finite $c$ we can see a sharp fermionic character appear that
broadens with increasing $c$ as shown in Fig.~\ref{fig:largec}.
The large time behavior is captured in a small window around
$\xi=0$. One can see that at any finite $c$, the region near zero
develops a strong fermionic character, thus indicating that
irrespective of the value of the coupling that we start with, the
model flows towards an infinitely repulsive model at large time, that
can be described in terms of free fermions. We also obtained this
result ``at'' $t=\infty$ at the beginning of this section.

\begin{figure}[htb]
  \centering \includegraphics[width=2.5in]{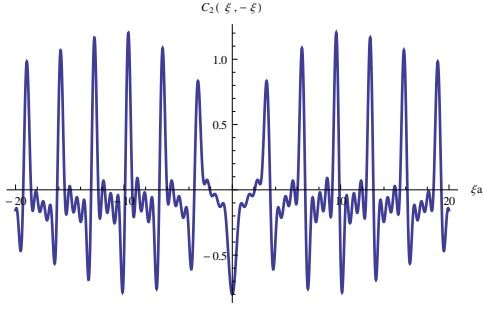}
  \caption{Normalized noise correlation function for five particles
    released for a Mott-like state for $c>0$ (from
    Ref.~\onlinecite{deeprl})}
  \label{fig:5prep}
\end{figure}

\begin{figure}[htb]
  \centering \includegraphics[width=2.5in]{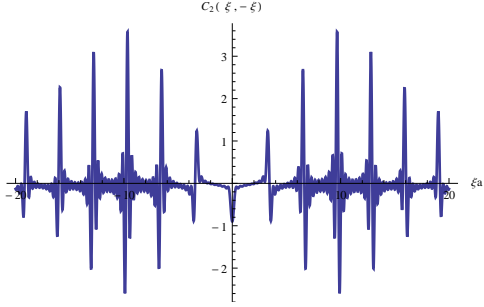}
  \caption{Normalized noise correlation function for ten particles
    released for a Mott-like state for $c>0$.}
  \label{fig:10prep}
\end{figure}

For higher particle number, we see ``interference fringes''
corresponding to the number of particles, that get narrower and more
numerous with an increase, memory of the initial lattice
state. However, the asymptotic fermionic character does not
disappear. Figures \ref{fig:5prep} and \ref{fig:10prep} show the noise
correlation function for five and ten particles respectively. The
large peaks are interspersed by smaller peaks and so on. This reflects
the character of the initial state.

\subsubsection{Quenching from a bound state}

\begin{figure}[htb]
  \centering \includegraphics[width=8cm]{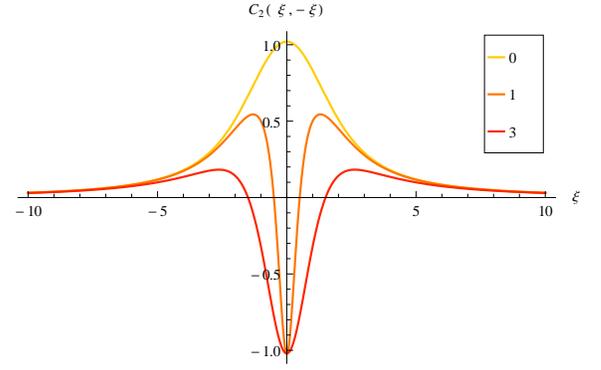}
  \caption{(Color online) Normalized noise correlation function for two particle quenched from a bound state into the repulsive regime.
  The legend indicates the values of $c$ that the state is quenched into. We start with $c_{0}\sigma^{2}=3,\;\sigma=1$,
  $c_{0}$ being the interaction strength of the initial state Hamiltonian.
  Again, we see the fermionic dip, but the rest of the structure is determined by the initial state.}
  \label{fig:boundquench}
\end{figure}

 In this brief section our initial state is the ground state of the attractive Lieb-Liniger Hamiltonian (with interaction strength $-c_{0}<0$.
 For two bosons, this take the form~\cite{lieblin},
 \begin{equation}
  \ket{\Psi_{\text{bound}}} = \int_{\vec{x}}e^{-c_{0}\abs{x_{1}-x_{2}} - \frac{x_{1}^{2}}{2\sigma^{2}}-\frac{x_{2}^{2}}{2\sigma^{2}}} b^{\dag}(x_{1})b^{\dag}(x_{2})\ket{0},
\end{equation}
and we quench it with a repulsive Hamiltonian.The long time noise correlations are displayed in Fig.~\ref{fig:boundquench}. We see that while the initial state
 correlations are preserved over most of the evolution, in the asymptotic long time limit the characteristic fermionic dip.
We expect similar effects for any number of bosons.

\subsection{Attractive interactions}
\label{sec:attasymp}
\begin{figure}[htb]
  \centering \includegraphics[width=6cm]{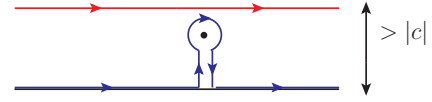}
  \caption{(Color online) Contribution from stationary phase and pole at large time
    in the attractive model. The blue (lower) contour represents the shifted
    contour.}
  \label{fig:attasymp}
\end{figure}

For the attractive case, since the contours of integration are spread
out in the imaginary direction, we have the contributions from the
poles in addition to the stationary phase contributions at large time.
The stationary phase contribution is picked up on the real line, but
as we move the contour, it stays pinned above the poles and we need to
include the residue obtained from going around them, leading to sum
over several terms.Fig.~\ref{fig:attasymp} shows an example of how
this works.

In Ref.~\onlinecite{deeprl}, a formula was provided for the asymptotic
state.  Here we give a more careful treatment by taking into account
that the fixed point of the approximation moves for terms that come
from a pole of the $S$-matrix. It is therefore necessary to first
shift the contours of integration, and then carry out the integral at
long time. We carry this out below.

Shifting a contour over a pole leads to an additional term from the
residue:
\begin{equation}
  \int_{\gamma_{2}}\frac{{\rm d}\lambda_{2}}{2\pi}\to \int_{\gamma^{\rm R}_{2}} \frac{{\rm d}\lambda_{2}}{2\pi} - i\mathcal{R}(\lambda_{2}\to\lambda_{1}+i\abs{c})
\end{equation}
where $\mathcal{R}(x)$ indicates that we evaluate the residue given by
the pole $x$. $\gamma_{j}$ indicates the original contour of
integration and $\gamma_{j}^{\rm R}$ indicates that integration is
carried out over the real axis.  Proceeding with the other variables
we end up with
\begin{multline} \label{eq:attasymp}
  \int_{\gamma_{1},\gamma_{2},\cdots,\gamma_{N}}\to \int_{\gamma^{\rm
      R}_{1}} \left[\int_{\gamma^{\rm R}_{2}} +
    i\mathcal{R}(\lambda_{2}\to\lambda_{1}+i\abs{c})\right]\\
  \times
  \left[\int_{\gamma^{\rm R}_{3}} + i\mathcal{R}(\lambda_{3}\to\lambda_{1}+i\abs{c}) + i\mathcal{R}(\lambda_{3}\to\lambda_{2}+i\abs{c})\right]\cdots\\
  \times\Big[\int_{\gamma^{\rm R}_{N}} +
  i\mathcal{R}(\lambda_{N}\to\lambda_{1}+i\abs{c}) +
  i\mathcal{R}(\lambda_{N}\to\lambda_{2}+i\abs{c}) + \\
  + \cdots + i\mathcal{R}(\lambda_{N}\to\lambda_{N-1}+i\abs{c}) \Big]
\end{multline}
The integrals can now be evaluated using the stationary phase
approximation. The correction produced by the above procedure does not
affect the qualitative features observed in Ref.~\onlinecite{deeprl}.

\subsubsection{Lattice initial state}
We now calculate the evolution of the density and the two body
correlation function in order to compare with the repulsive case.  We
will first study the two particle case. Although we have a finite time
expression for this case from which we can directly take a long time
limit, we will study the asymptotics using the above scheme for an
$N$-particle state, since we have an analytical expression to go with.
We get two terms, the first being the stationary phase contribution,
and is just like the repulsive case with $c\to-c$. The second is the
contribution from the pole. It contains the bound state contribution
which brings about another interesting feature of the attractive
case. While the asymptotic dynamics of the repulsive model is solely
dictated by the new variables $\xi_j\equiv\frac{y_j}{2t}$, and all the
time dependence of the wave function enters through this ``velocity''
variable, this is not the case in the attractive model. While it is
true that the system is naturally described in terms of $\xi$
variables, there still exists non-trivial time dependence.

First, we integrate out the $x$ dependence assuming an initial
lattice-like state. This gives,
\begin{multline}
  \label{eq:attasympnox}
  \ket{\Psi_{\text{latt}}(t)} = \int_y
  \sum_{\xi_j^*=\xi_j,\xi_i^*+ic,i<j}
  \prod_{i<j} S_{ij}(\xi_i^*-\xi_j^*)\prod_j \frac{(4\pi \sigma^2)^{\frac14}}{\sqrt{4\pi it}}\\
  \times e^{-(\sigma^2/2+it)(\xi_j^*)^2+i\xi_j^*(2t\xi_j+a(j-1))}
  b^\dag(y_j)\ket{0}.
\end{multline}
Defining $\phi(\xi,t)$ from $\ket{\Psi_{\text{latt}}(t)} = \int_y
\phi(\xi,t)\prod_j b^\dag(y_j)\ket{0}$, we have for the density
evolution under attractive interactions, $c<0$,
\begin{equation}
  \label{eq:densatt}
  \rho_{\text{latt}}^-(z;t) = \sum_{\{P\},j}\int_y \delta(y_j-z)\phi^*(\xi_P,t)\phi(\xi,t)
\end{equation}
We can show numerically (the expressions are a bit unwieldy to write
here), that asymptotically, the density shows the same Gaussian
profile that we expect from a uniformly diffusing gas, namely,
$e^{-\xi^2 \sigma^2}$.

\begin{figure}[h]
  \centering
  \includegraphics[width=8.5cm]{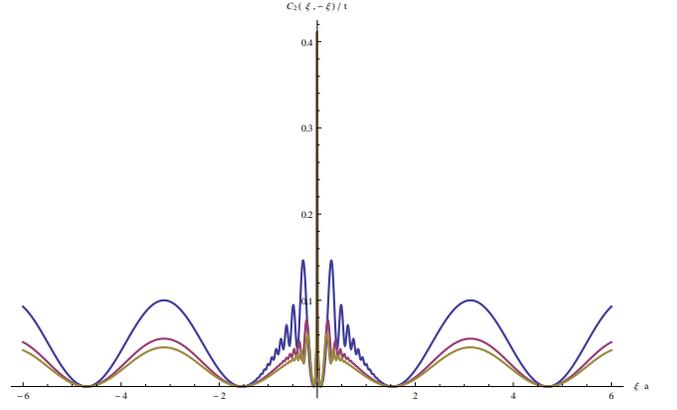}
  \caption{(Color online) Variation of $C_2$ for the attractive case
    with time. Note the growth of the central peak. At larger
    times, the correlations away from zero fall off. $ta^2=20,40,60$
    for blue (top), magenta (middle) and yellow (bottom) respectively.}
  \label{fig:c2atttime}
\end{figure}

With this, we can proceed to compute the noise correlation
function. The two particle case is easy, as there are no more
integrations to carry out. We get,
\begin{equation}\begin{split}
    \label{eq:noiseatt}
    \rho_{2\text{ latt}}^-(z,z';t) &= \sum_{\{P\},j,k}\int_y \delta(y_j-z)\delta(y_k-z')\phi^*(\xi_P,t)\phi(\xi,t)\\
    &= \lvert\phi_s(\xi_z,\xi_z')\rvert^2,
  \end{split}
\end{equation}
where $\phi_s$ is the symmetrized wavefunction.
Fig.\ref{fig:c2atttime} shows the normalized noise correlations for
different values of $t$.

\begin{figure}[h]
  \centering \includegraphics[width=8.5cm]{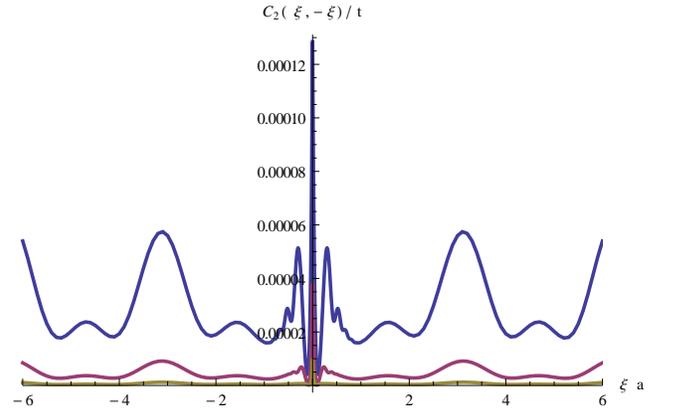}
  \caption{(Color online) $C_2(\xi,-\xi)$ for three particles in the
    attractive case plotted for three different times. At larger
    times, the correlations away from zero fall off. $ta^2=20,40,60$
    for blue (top), magenta (middle) and yellow (bottom) respectively. (from
    Ref.~\onlinecite{deeprl})}
  \label{fig:3patt}
\end{figure}

For more particles, we see interference fringes similar to the
repulsive case. We note that the central peak increases and
sharpens with time, indicating increasing contribution from bound
states to the correlations (see Fig.~\ref{fig:3patt} for an example).

\subsection{Starting with a condensate - attractive and repulsive
  interactions}
In this section, we study the evolution of the Bose gas after a quench
from an initial state where all the bosons are in a single level of a
harmonic trap. For $t<0$, the state is described by
\begin{equation}
  \ket{\Psi_{\text{cond}}} = \int_x \mathcal{S}_x \prod_j \frac{e^{-\frac{x_j^2}{\sigma^2}}}{(\pi\sigma^2)^{\frac14}} b^\dag(x_j)\ket{0}.
\end{equation}
Recall that in order to use the Yudson representation, the initial
state needs to be ordered. We can rewrite the above state as
\begin{multline}
  \ket{\Psi_{\text{cond}}} = \int_x\theta(x_1>\cdots>x_N) \\
  \times\mathcal{S}_x\prod_j
  \frac{e^{-\frac{x_j^2}{\sigma^2}}}{(\pi\sigma^2)^{\frac14}}
  b^\dag(x_j)\ket{0}
\end{multline}
where $\mathcal{S}$ is a symmetrizer. The time evolution can be
carried out via the Yudson representation, and again, we concentrate
on the asymptotics.  For the repulsive model, the stationary phase
contribution is all that appears, and we get
\begin{multline}
  \ket{\vec x} = \int_y \prod_{i<j} S^y_{ij}\left(\frac{y_i-y_j-x_i+x_j}{2t}\right)\\
  \prod_j \frac{1}{\sqrt{2\pi
      it}}e^{-i\frac{(y_j-x_j)^2}{4t}+i\frac{(y_j-x_j)^2}{2t}}
  b^\dag(y_j)\ket{0}.
\end{multline}
At large time $t$, we therefore have
\begin{multline}
  \ket{\Psi_{\text{cond}}(t)} = \int_{x,y} \theta(x_1>\cdots>x_N)\phi_2(x) I(y,x,t)\\
  \times \prod_j b^\dag(y_j)\ket{0}
\end{multline}
$\phi_2(x)$ is symmetric in $x$. $I(y,x,t)$ is symmetric in the $y$
but not in the $x$. Therefore we have to carry out the $x$ integration
over the wedge $x_1>\cdots>x_N$. This is not straightforward to carry
out. If $I(y,x,t)$ was also symmetric in $x$, then we can add the
other wedges to rebuild the full space in $x$. However, due to the
$S$-matrix factors, symmetrizing in $y$ does not automatically
symmetrize in $x$. The exponential factors on the other hand are
automatically symmetric in both variables if one of them is
symmetrized because their functional dependence is of the form
$f(y_j-x_j)$. It is however possible to make the $S$-matrix factors
approximately symmetric in $x$, and we will define what we mean by
approximately shortly. What is important is to obtain a $y_j-x_j$
dependence. As of now, the $S$-matrix that appears in the above
expression is
\begin{equation}
  S^y_{ij}\left(\frac{y_i-y_j-x_i+x_j}{2t}\right) = \frac{\frac{y_i-y_j-x_i+x_j}{2t} -ic\sgn(y_i-y_j)}{\frac{y_i-y_j-x_i+x_j}{2t} -ic}
\end{equation}
First, we can change $\sgn(y_i-y_j)$ to
$\sgn\left(\frac{y_i-y_j}{2t}\right)$ since $t>0$. Next, note that
asymptotically in time, the stationary phase contribution comes from
$\frac{y}{2t}\sim\mathcal{O}(1)$. However, since $x$ has finite
extent, at large enough time, $\frac{x}{2t}\sim 0$. We are therefore
justified in writing
$\sgn\left(\frac{y_i-x_i}{2t}-\frac{y_j-x_j}{2t}\right)$. The only
problem could arise when $y_i\sim y_j$. However, if this occurs, then
the $S$-matrix is approximately $\sgn(y_i-y_j)$ which is antisymmetric
in $ij$. With this prefactor the particles are effectively fermions,
and therefore at $y_i\sim y_j$, the wave-function has an approximate
node. At large time therefore, we do not have to be concerned with the
possibility of particles overlapping, and including the $x_i$ inside
the $\sgn$ function is valid. With this change the $S$-matrix
also becomes a function of $y_j-x_j$ and symmetrizing over $y$ one
automatically symmetrizes over $x$.

In short, we have established that the wave function asymptotically in
time can be made symmetric in $x$. This allows us to rebuild the full
space. We get
\begin{equation}
  \begin{split}
    \ket{\Psi_{\text{cond}}(t)} &= \int_{x,y} \sum_P\theta(x_P) \phi_2(x) I^s(y,x,t) \prod_jb^\dag(y_j)\ket{0} \\
    &= \int_{x,y} \phi_2(x) I^s(y,x,t) \prod_jb^\dag(y_j)\ket{0}
  \end{split}
\end{equation}
where the $s$ superscript indicates that we have established that
$I(y,x,t)$ is also symmetric in $x$. With this in mind, we can do away
with the ordering when we're integrating over the $x$ if we symmetrize
the initial state wave function and the final wave function.  Note
that when we calculate the expectation value of a physical observable,
the symmetry of the wavefunction is automatically enforced, and thus
taken care of automatically.

\begin{figure}[htb]
  \centering \includegraphics[width=8.5cm]{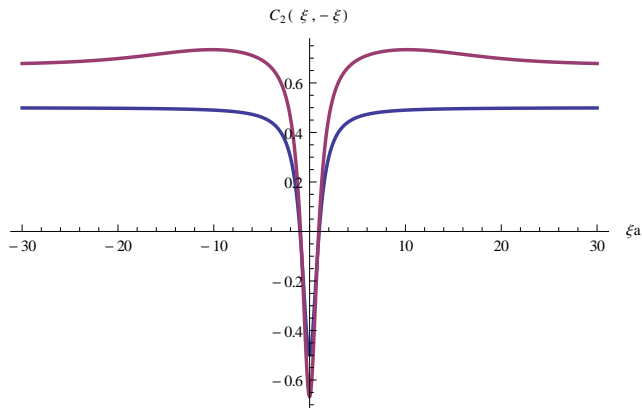}
  \caption{(Color online) $C_2(\xi,-\xi)$ for two (blue, bottom) and three (magenta, top)repulsive bosons
    starting from a condensate. Unlike the attractive case, there is
    no explicit time dependence asymptotically. $ca=3$ (from
    Ref.~\onlinecite{deeprl})}
  \label{fig:c2repcond23}
\end{figure}

\begin{figure}[htb]
  \centering \includegraphics[width=8.5cm]{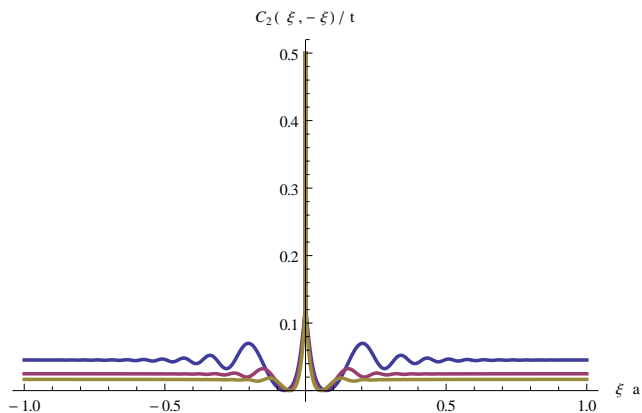}
  \caption{(Color online) Noise correlation for two attractive bosons starting from a
    condensate - as time increases from blue (top) to yellow (bottom), the central peak dominates.}
  \label{fig:c2attcondt}
\end{figure}

\begin{figure}[htb]
  \centering \includegraphics[width=8.5cm]{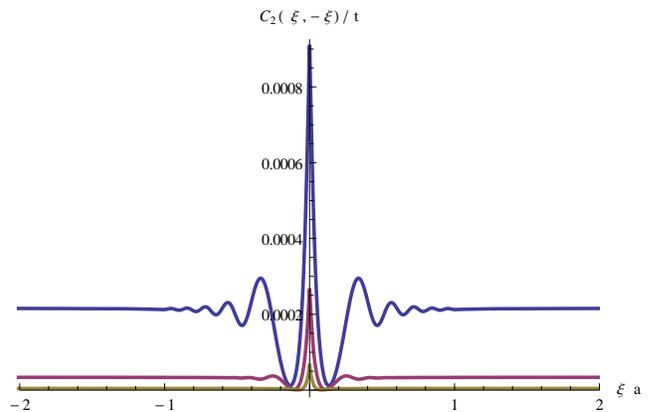}
  \caption{(Color online) $C_2(\xi,-\xi)$ for three attractive bosons starting from a
    condensate. Note that the side peak structure found in
    fig.~\ref{fig:3patt} is missing due to the initial condition. We
    show the evolution at three times. As time increases, the
    oscillations near the central peak die out. Times from top to
    bottom $t c^2=20,40,60$. (from Ref.~\onlinecite{deeprl})}
  \label{fig:c2attcond3}
\end{figure}

Recall that when we calculated the noise correlations of the repulsive
gas, in order to get an analytic expression for $N$ particles, we
considered the leading order term, i.e., the HBT term.  We did this by
showing that higher order crossings produced terms higher order in
$e^{-2ca}$ which we claimed was a small number. Now, however, $a=0$,
and although the calculation is essentially the same with our
approximate symmetrization, this simplification does not occur.  The
two and three particle results remain analytically calculable, but for
higher numbers, we have to resort to numerical
integration. Fig.~\ref{fig:c2repcond23} shows the noise correlation
for two and three repulsive bosons starting from a condensate. For
non-interacting particles, we expect a straight line $C_2=
\frac{1}{2}$. When repulsive interactions are turned on, we see the
characteristic fermionic dip develop. The plots for the attractive
Bose gas are shown in Figs.~\ref{fig:c2attcondt} and
\ref{fig:c2attcond3}. As expected from the non-interacting case the
oscillations arising from the interference of particles separated
spatially does not appear. The attractive however does show the
oscillations near the central peak that are also visible in the case
when we start from a lattice-like state. It is interesting to note
that for three particles we do not see any additional structure
develop in the attractive case.


\section{Conclusions and the dynamic RG hypothesis}
\label{sec:concl}
We have shown that the Yudson contour integral representation for
arbitrary states can indeed be used to understand aspects of the quench
dynamics of the Lieb-Liniger model, and obtain the asymptotic wave functions exactly.
The
representation overcomes some of the major difficulties involved in
using the Bethe-Ansatz to study the dynamics of some integrable
systems by automatically accounting for complicated states in the spectrum.

We see some interesting dynamical effects at long times.  The infinite
time limit of the repulsive model corresponds to particles evolving
with a free fermionic Hamiltonian. It retains, however, memory of the
initial state and
therefore is not a thermal state. The correlation functions approach
that of hard core bosons at long time indicating a dynamical increase
in interaction strength. The attractive model also shows a dynamic
strengthening of the interaction and the long time limit is dominated
by a multiparticle bound state. This of course does not mean that it condenses. In fact the
state diffuses over time, but remains strongly correlated.

We may interpret our results in terms of a ``dynamic RG'' in time.
The asymptotic evolutions of the model both for $c>0$ and for $c<0$
are given by the Hamiltonians $H^*_{\pm}$ with $c\to\pm \infty$
respectively. Accepting the RG logic behind the conjecture one would
expect that there would be basins of attraction around the
Lieb-Liniger Hamiltonian with models whose long time evolution would
bring them close to the "dynamic fixed points" $H^*_{\pm}$. One such
Hamiltonian would have short range potentials replacing the
$\delta$-function interaction that renders the Lieb-Liniger model
integrable. Perhaps, lattice models could be also found in this basin
whose time asymptotics would be close, in the repulsive case, to that
given by a free fermionic model on the lattice. Clearly, as discussed
earlier, the Bose-Hubbard model is not such a model since it has a
lattice symmetry that is not present in the Lieb-Liniger
model. This could be however overcome by adding such terms as the next
nearest hopping or interactions that break this symmetry, or as shown in Appendix~\ref{sec:bhtebd},
with an appropriate choice of initial state.

\begin{figure}[htb]
  \centering
  \includegraphics[width=8.5cm]{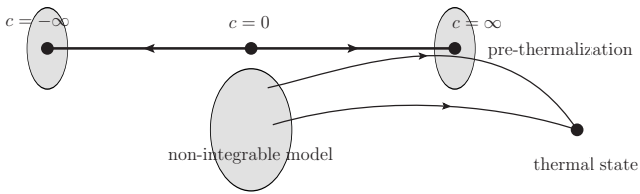}
  \caption{Schematic showing pre-thermalization of states in a
    non-integrable model}
  \label{fig:pretherm}
\end{figure}

We have to emphasize, however, that as these models are not integrable,
we do not expect that they would actually flow to $H^*_{\pm}$. Instead, starting close enough in the ``basin'', they would flow close
to $H^*_{\pm} $ and spend much time in its neighborhood, eventually
evolving into another, thermal state.  We thus conjecture that away
from integrability, a system would approach the corresponding
non-thermal equilibrium, where the dynamics will slow down leading to
a ``prethermal'' state~\cite{berges}. Fig.~\ref{fig:pretherm} shows a
schematic of this.  Such prethermalization behavior has indeed been
observed in lattice models~\cite{kollath}. The system is expected to
eventually find a thermal state.  It is therefore of interest to
characterize different ways of breaking integrability to see when a
system is ``too far'' from integrability to see this effect and in
what regimes a system can be considered as close to
integrability. For a review and background on this subject,
see Ref.~\onlinecite{polkovnikovrmp}.

Further, the flow diagram in Fig.~\ref{fig:pretherm} might have another axis that represents initial states.
Studying the Bose-Hubbard model  shows an interesting initial state dependence.
Whereas the sign of the interaction does not affect the quench dynamics,
the asymptotic state depends strongly on the initial state, with a lattice-like
state leading to fermionization, and a condensate-like state retaining bosonic
correlations. The strong dependence on the initial state in the quench dynamics is evident from
eq.~\eqref{eq:quenchoverlaps} and is subject of much debate, in particular as relating to the Eigenstate
Thermalization Hypothesis~\cite{rigol2,srednickitherm,deutsch}.

This work also opens up several new questions. It provides a prediction for experiments that can be carried out in the context of continuum cold
atom systems (though the experiments we are aware of are carried out on the lattice and therefore described by the Bose-Hubbard model)
Theoretically, while the
representation is provable mathematically, further investigation is
required to understand, physically, how it achieves the tedious sum
over eigenstates, while automatically accounting for the details of
the spectrum. This would allow us to extend the approach to other models
with a more complicated $S$-matrix structure.
It would also be useful to tie this approach to other means of calculating
overlaps in the Algebraic Bethe Ansatz, i.e., the form-factor approach.
The representation can essentially be thought of as a different way
of writing the identity operator. From that standpoint, it could serve
as a new way of evaluating correlation functions using the Bethe
Ansatz. We are currently studying generalizations of this
approach to other models that can be realized in optical lattices.

\section{Acknowledgments}
\label{sec:ack}
We are grateful to G.~Goldstein for very
useful discussions. This work was supported by NSF grant DMR 1006684.

\appendix
\section{Quenching the Bose-Hubbard model}
\label{sec:bhtebd}
We compare the results obtained in Section~\ref{sec:2pquench} with
those from the lattice version of the Lieb-Liniger model - the
Bose-Hubbard model,
\begin{equation}
  H_{\text{BH}} = \sum_{i}\left[\left(tb^{\dag}_{i}b_{i+1}+\text{h.c.}\right) + Un_{i}(n_{i}-1)\right]
\end{equation}
It describe bosons $b$ hopping on a 1$d$ lattice with on-site
interaction $U$ and is non-integrable since it allows multiparticle
interactions on the same site. It has been extensively studied in many
contexts and much is known about its equilibrium properties (see e.g.,
Ref.~\onlinecite{kuhner}). For $0<U/t\ll 1$, the model is a
superfluid, and for $U/t\gg 1$ it is a Mott insulator.  For negative
$U$, the model is attractive and the ground state is a Bose
condensate.  A \emph{non-equilibrium phase diagram} of the
Bose-Hubbard model is given in Ref.~\onlinecite{kollath}.

We study here the two boson quench dynamics and contrast it with the
corresponding dynamics of the Lieb-Liniger model. Contrary to what one
may expect, the introduction of the lattice modifies the dynamics in
an essential way even at long times and distances.  The calculations
of density correlations as a function of time after a sudden quench
have been carried out using the \emph{Algorithms and Libraries for
  Physics Simulations} (ALPS) code~\cite{alpsurl,alps1,alps2} and the
\emph{Open source TEBD} package~\cite{tebdurl} after making the
necessary modifications to accommodate the initial states we are
interested in. Our results confirm some results obtained in
Ref.~\onlinecite{lahinietal}.

\begin{figure}[htb]
  \centering \includegraphics[width=3in]{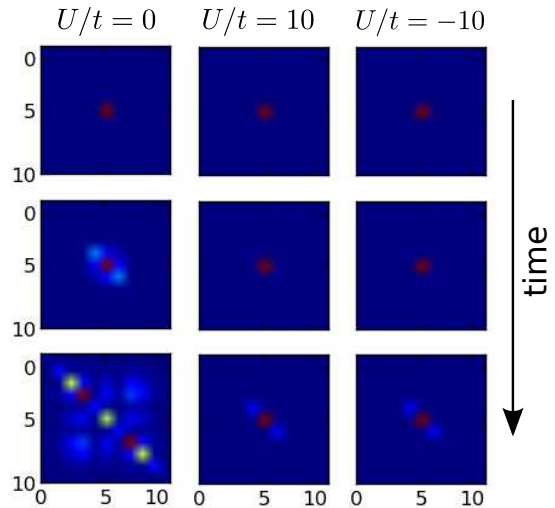}
  \caption{(Color online) Time evolution of the correlation matrix
    after a sudden quench from a state containing two bosons on the
    same site. The values increase from blue (0) to red. The correlations remain strong
  in the center indicating strong bunching.}
  \label{fig:tebdcond}
\end{figure}

\begin{figure}[htb]
  \centering \includegraphics[width=3in]{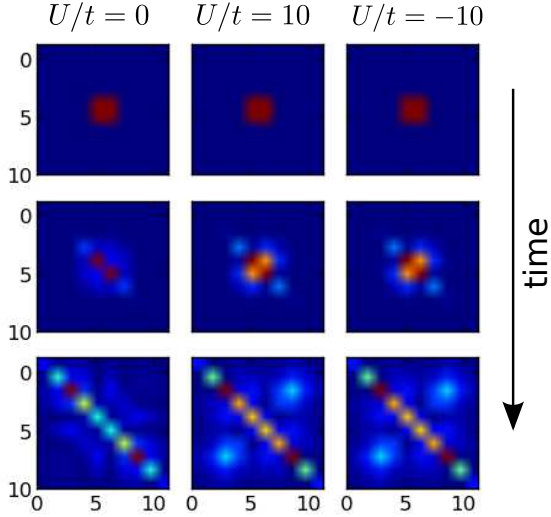}
  \caption{(Color online) Time evolution of the correlation matrix
    after a sudden quench from a state containing bosons on two
    neighboring sites.  The values increase from blue (0) to red. The off diagonal
    correlations indicate anti-bunching, as can be seen from free fermion evolution in Fig.~\ref{fig:fermionlatt}.}
  \label{fig:tebdlatt}
\end{figure}

\begin{figure}[htb]
  \centering \includegraphics[width=1.2in]{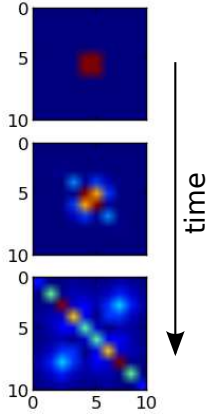}
  \caption{(Color online) Time evolution of free fermions on a lattice. Notice how off diagonal
    correlations develop. The values
    increase from blue (0) to red.}
  \label{fig:fermionlatt}
\end{figure}

In Fig.~\ref{fig:tebdcond} we show the time evolution of the
correlation matrix defined as $\langle n_{i}n_{j}\rangle$ after a
sudden quench from an initial state $b^{\dag}_{0}b^{\dag}_{0}\ket{0}$,
and in Fig.~\ref{fig:tebdlatt} the evolution from initial state
$b^{\dag}_{0}b^{\dag}_{1}\ket{0}$.  We quench into the interacting
regime, where $|U|/t = 10$.  There are a couple of interesting
features: (1) Unlike the situation in the Lieb-Liniger model where the
bunching or anti-bunching effect is independent of the initial state,
here, quenching a lattice-like state leads to anti-bunching
Fig.~\ref{fig:tebdlatt}, while quenching a condensate-like state leads
to bunching Fig.~\ref{fig:tebdcond}. It is also interesting to compare
the anti-bunching evolution of the bosons with the evolution of
free fermions in Fig.~\ref{fig:fermionlatt}.
(2) The sign of the interaction plays no role in the evolution in the
Bose-Hubbard model, as seen from either figures.  This is unlike the
situation in the continuum model where for repulsive interactions
anti-bunching (fermionization) occurs independently of the initial
state, while bunching will take place for attractive interactions.

This non-dependence on the sign of the interaction is due to a
particle-hole symmetry that is present on the lattice, but not in the
continuum. The 1$d$ lattice, being bipartite, allows the the
transformation $b_j \to e^{i\pi j}b_j$, with$j$ the site index, under
which the hopping terms pick up a minus sign, while the on-site
interaction terms are unaffected, thus $U/t\to -U/t$. In terms of
$\mathcal{U}$ the corresponding unitary operator we have,
\begin{equation}
  \mathcal{U} H_{\text{BH}}(t,-U)\mathcal{U}^{\dag} = -H_{\text{BH}}\,(t,U).
\end{equation}
Denoting eigenstates and eigenvalues of $H_{\text{BH}}(t,U)\equiv
H_{\text{BH}}$ by $\ket{m}$ and $\epsilon_m$ and the corresponding
eigenstates and eigenvalues of $H_{\text{BH}}(t,-U) \equiv
\tilde{H}_{\text{BH}}$ by $\ket{\tilde{m}}$ and $\tilde{\epsilon}_{m}$
we can relate the states by: $\ket{\tilde{m}} = \mathcal{U}\ket{m}$,
and the eigenvalues by: $\tilde{\epsilon}_{m}=-\epsilon_{m}$.  The
time evolution of an operator $\mathcal{O}$ under the action of
$H_{\text{BH}}(t,U)$ from an initial state $\ket{\Psi_{0}}$
\begin{equation}
  \langle\mathcal{O}(t)\rangle_{H} = \sum_{m,m'} \langle\Psi_{0}|m'\rangle
  \langle m|\Psi_{0}\rangle\bra{m'}\mathcal{O}\ket{m}e^{-i(\epsilon_{m}-\epsilon_{m'})t}
\end{equation}
Under $\tilde{H}_{\text{BH}}$,
\begin{equation}
  \begin{split}
    &\langle\mathcal{O}(t)\rangle_{\tilde{H}} =
    \sum_{\tilde{m},\tilde{m}'}
    \langle\Psi_{0}|\tilde{m}'\rangle\langle \tilde{m}|\Psi_{0}\rangle
    \bra{\tilde{m}'}\mathcal{O}\ket{\tilde{m}}e^{-i(\tilde{\epsilon}_{m}- \tilde{\epsilon_{m'}})t}    \\
    &= \sum_{m,m'}
    \bra{\Psi_{0}}\mathcal{U}\ket{m'}\bra{m}\mathcal{U}^{\dag}\ket{\Psi_{0}}
    \bra{m'}\mathcal{U}^{\dag}\mathcal{O}\mathcal{U}\ket{m}e^{i(\epsilon_{m}-\epsilon_{m'})t}
  \end{split}
\end{equation}
Both initial state we considered, $\ket{\Psi_{\text{latt}}} =
b_{0}^{\dag}b_{1}^{\dag}\ket{0}$ and
$\ket{\Psi_{\text{cond}}}=(b^{\dag}_{0})^{2}\ket{0}$ are simply
transformed, $\mathcal{U}\ket{\Psi_{0}} = \pm\ket{\Psi_{0}}$ and as
they occur twice in the overlaps the transformation leaves no effect.
Similarly the operators we have considered (density-density
correlations) are bilinear in the site operators and are therefore not
affected, $\mathcal{U}^{\dag}\mathcal{O}\mathcal{U} =
\mathcal{O}$. This gives
\begin{equation}
  \langle\mathcal{O}(t)\rangle_{\tilde{H}} = \sum_{m,m'}
  \langle\Psi_{0}\ket{m'}\langle m\ket{\Psi_{0}}
  \bra{m'}\mathcal{O}\ket{m}e^{i(\epsilon_{m}-\epsilon_{m'})t}
\end{equation}
Next, we note that the Bose-Hubbard Hamiltonian is invariant under
time reversal, i.e., $[H_{\text{BH}},\mathcal{T}]=0$ where
$\mathcal{T}$ is the anti-unitary time reversal operator:
\begin{equation}
  \mathcal{T}t\mathcal{T}^{-1}=-t,\qquad \mathcal{T}i\mathcal{T}^{-1}=-i.
\end{equation}
Applying the time reversal operator to the expectation value above, we
get
\begin{equation}
  \begin{split}
    &\langle\mathcal{O}(t)\rangle_{\tilde{H}} =   \mathcal{T}\langle\mathcal{O}(t)\rangle_{\tilde{H}}\mathcal{T}^{-1} \\
    & = \sum_{m,m'} \langle\Psi_{0}\ket{m'}^{*}\langle
    m\ket{\Psi_{0}}^{*}
    \bra{m'}\mathcal{O}\ket{m}^{*}e^{i(\epsilon_{m}-\epsilon_{m'})t}\\
    & = \sum_{m,m'} \langle \Psi_{0}\ket{m}\langle m'\ket{\Psi_{0}}
    \bra{m}\mathcal{O}\ket{m'}e^{i(\epsilon_{m}-\epsilon_{m'})t}\\
    &= \sum_{m,m'} \langle \Psi_{0}\ket{m'}\langle m\ket{\Psi_{0}}
    \bra{m'}\mathcal{O}\ket{m}e^{-i(\epsilon_{m}-\epsilon_{m'})t}\\
    &= \langle\mathcal{O}(t)\rangle_{H}
  \end{split}
\end{equation}
thus indeed,  the time evolution looks the
same for both signs of the interaction.  Note that  with initial states or operators
that are not invariant (up to
a sign) under the transformation $\mathcal{U}$, we should see a
difference in the time evolution of the attractive and repulsive
models.

A similar symmetry exists in the XXZ model or the Hubbard model in
1$d$ (or higher dimensional bipartite lattices). For the magnet, the

sign of the anisotropy $\Delta$ leads to either ferromagnetic or
antiferromagnetic ground states for negative or positive
anisotropy. However, it does not influence the quench
dynamics~\cite{barmettler}, as can be seen from arguments like the
above. Similarly in the Hubbard model, the quench dynamics is
unaffected by the change of sign of $U$~\cite{schneider}.

\bibliography{bosegas}

\end{document}